\documentclass[a4paper]{article}

\usepackage[margin=1in]{geometry} % full-width

\usepackage{amsmath}
\usepackage{amsthm}
\usepackage{amssymb}
\usepackage[table,xcdraw]{xcolor}
\usepackage{longtable}
\usepackage{float}
\usepackage{wasysym}
\usepackage[utf8]{inputenc}
\usepackage{hyperref}
\usepackage{oplotsymbl}
\usepackage{graphicx, color}
\usepackage{mathrsfs} % for \mathscr command

\usepackage{lipsum}
\usepackage[T1]{fontenc}
\usepackage{comment}
\usepackage{multirow}
\usepackage{longtable}
\usepackage{booktabs}
\usepackage{lscape}
\usepackage{url}
\usepackage{subcaption}

\title{DApps Ecosystems: Mapping the Network Structure of Smart Contract Interactions}

\author{\small Sabrina Aufiero$^{1, *}$, Giacomo Ibba$^2$, Silvia Bartolucci$^1$, Giuseppe Destefanis$^3$, Rumyana Neykova$^3$, Marco Ortu$^2$ \\  }

\date{ \small $^1$ Dept. of Computer Science, University College London (UK) \\  $^2$ Dept. of Business and Economic Sciences, University of Cagliari (Italy) \\ $^3$ Dept. of Computer Science, Brunel University (UK) \\ $^\star$ Corresponding author: \url{sabrina.aufiero.22@ucl.ac.uk}
}

\begin{document}

\flushbottom
\maketitle

\begin{abstract}
In recent years, decentralized applications (dApps) built on blockchain platforms such as Ethereum and coded in languages such as Solidity, have gained attention for their potential to disrupt traditional centralized systems.  Despite their rapid adoption, limited research has been conducted to understand the underlying code structure of these applications. In particular, each dApp is composed of multiple smart contracts, each containing a number of functions that can be called to trigger a specific event, e.g., a token transfer. In this paper, we  reconstruct and analyse the network of contracts and functions calls within the dApp, which is helpful to unveil vulnerabilities that can be exploited by malicious attackers. We show how decentralization is architecturally implemented, identifying common development patterns and anomalies that could influence the system's robustness and efficiency. We find a consistent network structure characterized by modular, self-sufficient contracts and a complex web of function interactions, indicating common coding practices across the blockchain community. Critically, a small number of key functions within each dApp play a pivotal role in maintaining network connectivity, making them potential targets for cyber attacks and highlighting the need for robust security measures.
\end{abstract}

\section{Introduction} \label{sec:Introduction}

\begin{figure}[]
    \centering
    \includegraphics[width=0.7\textwidth]{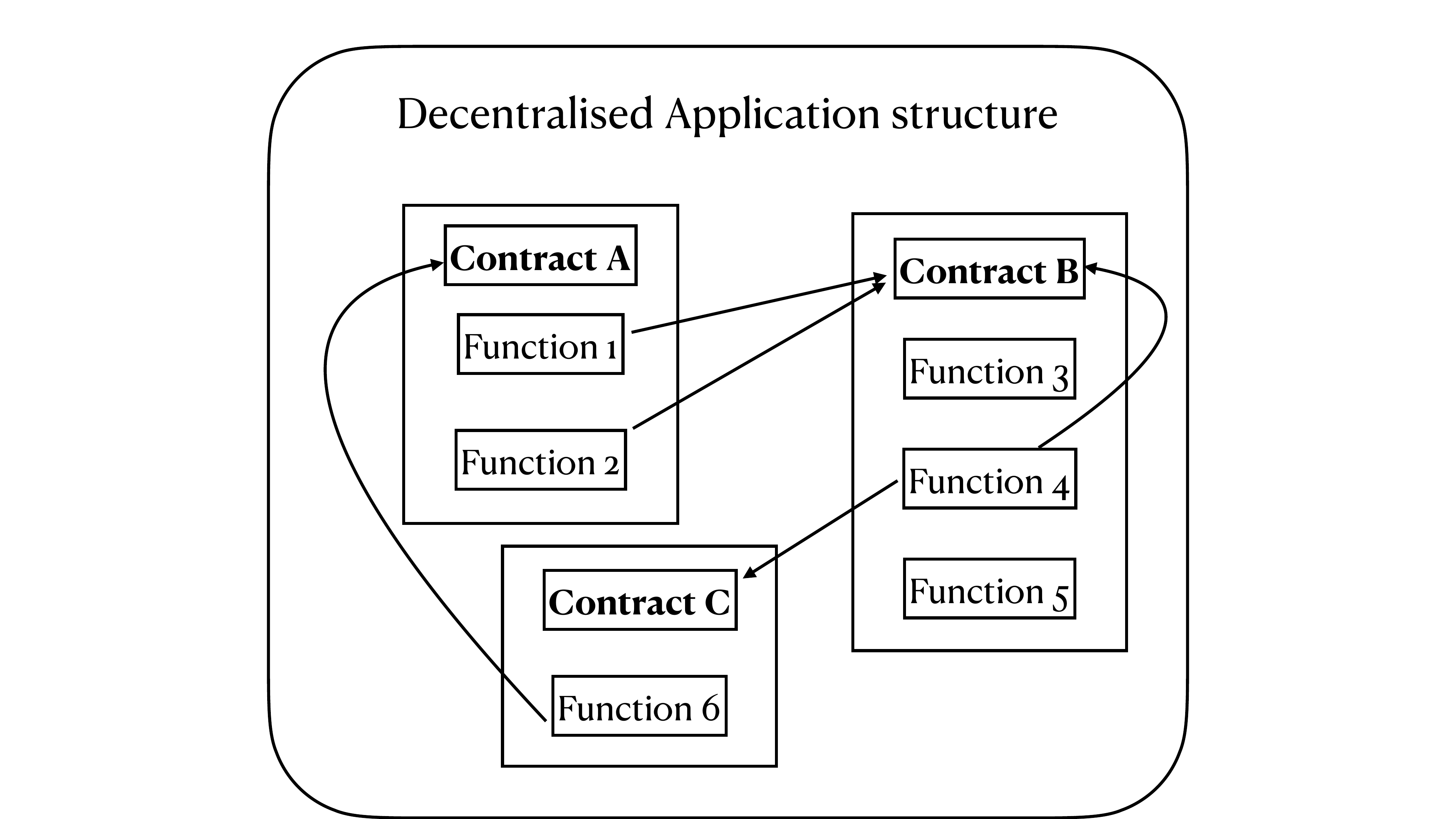}
    \caption{Scheme of a dApp structure. The dApp is composed of $3$ contracts, each with a varying number of functions. Contracts can interact between them via function calls (black arrows).}
    \label{fig:scheme}
\end{figure}

In recent years, the Total Value Locked (TVL) in decentralised finance platforms and crypto protocols has reached $44$ bUSD, with over $82$ million wallets and active users worldwide.
Since the launch of Bitcoin, blockchains and decentralised platforms have evolved to enable new functionalities and use cases beyond digital currency. 
These functionalities are embedded in smart contracts, a digital agreement written in code, stored on a blockchain, and executed automatically without intermediaries \cite{antonopoulosETH}. Smart contracts benefit from the blockchain’s security and transparency, providing users with a way to enforce agreements and streamline processes, and they are decentralized so they cannot be changed or tampered with once they are deployed. These terms can be as simple as making a single payment, or as complex as a multi-step process with many participants and data point requirements. Once deployed, anyone with access to the blockchain can invoke and interact with the smart contract.
Multiple contracts can be linked together to form a more sophisticated application, with different purposes and functionalities. Notable use cases include facilitating financial transactions or gaming interactions, enabling players to own and trade in-game assets, participate in competitions, and earn rewards in the form of tokens. These decentralized applications operating on blockchain systems, also widely called dApps, enhance many traditional industries and services, and are not run or controlled by a single central authority or trusted organisation. 
DApps are developed in most cases in an open-source fashion on Github: they are composed of a collection of {\em contracts}, each containing multiple {\em functions} that can be called by the same or different contracts, one or multiple times, depending on the task performed by the user (see Fig. \ref{fig:scheme}).

\begin{figure}
    \centering
    \includegraphics[width=0.7\textwidth]{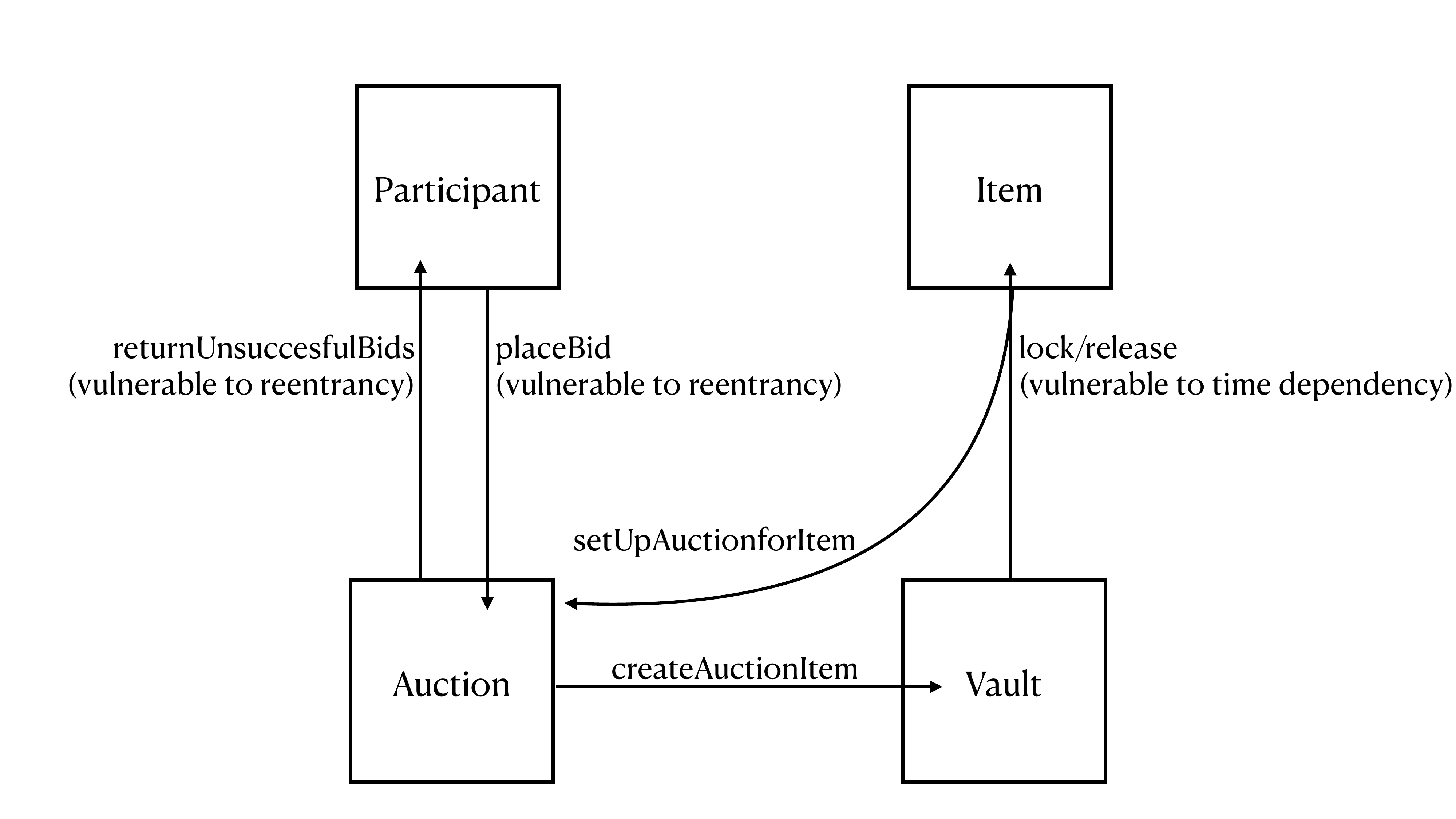}
    \caption{Scheme of a toy dApp example vulnerable to re-entrancy and time dependency attacks.}
    \label{fig:auction scheme}
\end{figure}

While dApps offer various advanced features like transparency and community collaboration, they may not be completely immune to security breaches or hacking attempts, and that is why a robust analytical framework is needed to study their complexity and vulnerabilities. 
One of the issues that users are usually concerned about is technical vulnerabilities, and while there are usually strong measures and smart contract bug bounty programs in place to address these issues \cite{qian2022smart}, smart contracts can still be exploited by malicious actors, leading to financial loss or unintended consequences. As opposed to traditional finance, developers of Decentralised Finance (DeFi) projects often lack financial experience, and cyber security is an afterthought in a hasty development process \cite{oosthoek2021flash}. Moreover, the hosting of smart contract code publicly on Github further enables an attacker’s opportunity to locate vulnerabilities quickly and efficiently. Exploiting DeFi projects currently is a low-risk high-reward opportunity to malicious actors. For example in the past years, the Decentralized Finance sector has experienced thousands of attacks causing the loss of millions of dollars locked in protocols \cite{gudgeon2020decentralized}, which could be dramatically reduced by actively monitoring and fixing security threats related to bugs in the code.
Another major challenge with dApps is scalability \cite{wu2021first}. Some blockchains have limitations in terms of processing speed and capacity, which can result in slower transaction times and higher costs. Scalability becomes a major concern especially when the number of users and transactions increases. 
Often, these vulnerabilities are directly linked to the way in which contracts and functions interact.
\\
Let us consider a toy dApp example illustrating technical vulnerabilities. In Fig. \ref{fig:auction scheme}, we present a simplified dApp responsible for managing the buying and selling of items during an auction. There are four contracts involved: \texttt{Auction}, \texttt{Item}, \texttt{Participant}, and \texttt{Vault}. The  \texttt{Auction} contract handles auction management, registering buyers, sellers, and items. The \texttt{Item} contract manages auction items and updates related to offers on an item. \texttt{Participant} manages the participants, allowing them to enrol as buyers or sellers and keeps track of their respective bids. Lastly,  \texttt{Vault} is responsible for safeguarding items and preventing them from being assigned to anyone before the auction's end. The vulnerable functions in this context are \texttt{lock, release, placeBid}, and \texttt{returnUnsuccessfulBids} \cite{he2020smart}. \texttt{placeBid} and \texttt{returnUnsuccessfulBids} are susceptible to reentrancy attacks. Consequently, an attacker, by reentering the function's code multiple times, could illicitly withdraw the funds related to the bids (funds meant to be collected by the auctioneer). On the other hand, the lock and release functions are vulnerable to time dependency issues, potentially causing items to be locked indefinitely or released prematurely before the end of the auction. As a result, participants may purchase items, which they will never gain ownership of, or items could be assigned to participants who did not place the highest bid during the auction.
The reentrancy attack has been exploited in the famous DAO Hack, where an attacker was able to call the function SplitDAO recursively, transferring 
 $\sim 50$ mUSD in its account \cite{zhao2017dao}.
 Note also that each function call has an associated computational cost to run and execute the code, and a fee paid to the network for validation (i.e., the so-called gas fees in Ethereum). In terms of scalability, a more complex call structure corresponds to higher computational costs and fees to execute a given action.
 
This research, leveraging complex network analysis tools, intends to study the dApps' complexity and characterize development practices. Our aim is to provide an understanding of the code structure underlying dApps deployed on different blockchains. This offers insights on architectural choices, vulnerabilities detection, and future development directions of decentralized systems. In particular:

\begin{enumerate}
    \item We identify critical components of the dApp, more likely to be susceptible to technical vulnerabilities.
    \item We find common practices concerning the implemented architecture of interactions between functions and contracts across blockchain ecosystems and development teams.
\end{enumerate}

These results can be used to inform development guidelines and active monitoring, ensuring a safer ecosystem for end-users. Indeed, recently regulators have been looking more closely at ways to tackle and minimise malicious activities in crypto markets \cite{board2023financial}, and businesses have joined forces to put forward best practices to ensure an increased trust in the technology and support adoption.

In the following, we will focus our analysis on $66$ dApps written in Solidity, a widely adopted, high-level programming language specifically designed for writing smart contracts on blockchain platforms (e.g., Ethereum). In Sec. \ref{sec:Related Works}, we discuss related and complementary literature. The dataset and methodology will be discussed in details in Sec. \ref{sec:Methodology}. Finally, we will present the main results in Sec. \ref{sec:Results}, and we will discuss them in Sec. \ref{sec:discussion} pointing to future research directions.

\section{Related Works} \label{sec:Related Works}

Complex systems approaches have shed light on the users' interactions \cite{farmer2012complex}, platform's growth, evolution and resilience \cite{linkov2019fundamental}, and market dynamics of crypto ecosystems \cite{soloviev2019complex}.  In the context of blockchain open-source development, the interplay between developers' team interactions on Github and market behavior of associated cryptocurrencies has been explored \cite{lucchini2020code, bartolucci2020butterfly}, highlighting a strong inter-dependence between the code development and assets' valuation.
More specifically, complex networks approaches have also been used to analyse blockchain transactions and addresses interactions, to characterise users' behaviour \cite{ferretti2020ethereum}, track malicious activities \cite{la2023game}, and identify links with cryptocurrency price dynamics \cite{bovet2023evolving}.

Within the software engineering community, complex network tools have been increasingly used to analyse characteristics of the underlying code structure. The most common approach assumes that software modules are represented as nodes, while relations among them correspond to edges. Other software artifacts, but also people involved in the software development process, have been considered as nodes leading to different kinds of networks. Modeling software systems as networks enabled a graph-based treatment and analysis with the goal of investigating several properties, such as scale-freeness, and the presence of small-world phenomena \cite{louridas2008power, potanin2005scale, kleinberg2000small, valverde2003hierarchical}. Object-oriented designs in particular, can be naturally represented as graphs \cite{chaikalis2014forecasting}. Software is built up out of many interacting units and subsystems at many levels of granularity (subroutines, classes, source files, libraries, etc.), and the interactions and collaborations of those pieces can be used to define networks or graphs that form a description of a system \cite{myers2003software}.  In addition, software code remains predominantly a handmade product, produced by human developers, and as such, it is prone to error. The result of a developer error can be directly translated into faults in code and as the world demands ever larger and more complex software systems, controlling faults in code becomes more difficult but increasingly necessary. Understanding fault insertion and fault fixing is crucial to enabling the effective reduction of faults in software systems \cite{ortu2023fault}. \\
In the context of blockchain systems, understanding the network interactions within and among smart contracts could provide new perspectives on system vulnerabilities and operational efficiencies \cite{zou2019smart}. 
Recently, researchers have started looking at defining rules and metrics to evaluate smart contract code specifically,  within the realm of the so-called blockchain-oriented software engineering research \cite{destefanis2018smart}.
A number of tools have been developed to analyse code and detect known and typical vulnerabilities, such as {\em Mythril} and {\em Osiris} for a smart contract static analysis, {\em Maian} that detects smart contract vulnerabilities by using dynamic analysis, and {\em Gasper} used to monitor the gas consumption of smart contracts \cite{qian2022smart}. Preliminary classifications of typical smart contracts vulnerabilities, such as re-entrancy, computational complexity and overflow have also been conducted \cite{he2020smart,huang2019smart}.
In a recent work, Ibba et al. \cite{ibba2018preliminary} examines software metrics in dApps to analyse their structural and behavioral characteristics as they grow in complexity. However, to the best of our knowledge, there is limited research on the applicability of complex network theory to the analysis of smart contracts' and dApps' code structure. 

In this work, we aim to bridge this gap by proposing a {\em complex networks driven software engineering} approach. The dApp's underlying code structure is represented and analysed as a network, whose nodes are functions and contracts, and links represent the strength of the interactions between them.
\section{Dataset and Methods} \label{sec:Methodology}
Solidity is a high-level programming language specifically designed for writing smart contracts on blockchain platforms \cite{antonopoulosETH}. It incorporates elements of pre-existing languages such as JavaScript and Python, but is tailored to the requirements of blockchain development. One of its standout features is its contract-oriented design, which allows for the construction of modular and reusable code structures. This enables developers to create decentralised applications, capable of reproducing complex real-world processes. Analysing Solidity smart contracts is of paramount importance -- given its widespread adoption --  to assess two critical aspects: platforms' security and robustness of the structural design of dApps. Security vulnerabilities in smart contracts can be dangerous \cite{destefanis2018smart}, given the immutable nature of blockchain, while the study of Solidity contracts and functions interactions allows to investigate the architecture and operational logic underlying dApps. Indeed, contracts contain the rules and functions that dictate the behaviour of a dApp, making their analysis crucial for understanding how these decentralised systems function. In the following sections we introduce the main steps to gather the data -- together with summary statistics and qualitative analysis of the data -- and construct contract and function networks.
\subsection{Data extraction and parsing}
In this work, we focus on a dataset composed of dApps mainly supported by the \href{https://ethereum.org/en/dapps/}{Ethereum} blockchain, but including also examples from other blockchains as \href{https://academy.binance.com/en/glossary/decentralized-application}{Binance}, \href{https://www.optimism.io/apps/all} {Optimism},  \href{https://polygon.technology/ecosystem}{Polygon}, \href{https://astar.network}{Astar}, \href{https://shiden.astar.network}{Shiden} and \href{https://www.thundercore.com}{ThunderCore}. 
The data on the underlying smart contract code is obtained from the Github repository of each dApp. For each dApp, the associated smart contracts code is broken down into relevant sub-components (e.g., libraries, functions, etc.) using an {\em ad hoc} tool specifically built to recognize these sub-parts in Solidity contracts \cite{ibba2023mindthedapp}. More specifically, we use the tool {\em MindTheDApp}, designed for the structural analysis of dApps built with Solidity contracts \cite{ibba2023mindthedapp}. The tool uses \href{https://www.antlr.org/}{ANTLR4} \cite{parr2013definitive} to traverse the Abstract Syntax Tree (AST) --  a tree representation of the abstract syntactic structure of the source code -- of Solidity contracts.  ANTLR4 works by accepting grammar rules to automatically produce both a lexer and a parser. The lexer first breaks down the input Solidity code into tokens, eliminating unnecessary elements such as whitespaces and comments. These tokens are then processed by the parser to form an Abstract Syntax Tree (AST), which organizes the code into a hierarchical structure useful for the analysis.

\subsection{Dataset features}
For each dApp the tool gives as output a CSV file, containing information regarding functions invoking contracts, allowing for advanced network analysis. The dataset obtained from our parser comprises for each dApp the information on the File in which the contract is defined, on the Source Function, i.e. the function calling a target contract (for example Function 1 in Fig. \ref{fig:scheme}), the Source and Target Contract, respectively the contract which the function belongs to and the contract the function is called by (Contract A and B in Fig. \ref{fig:scheme}). In Tab. \ref{tab:output parser} we provide an example of the parser output for {\em Aave} (category: Ethereum - DeFi; balance: $\$108.85B $; ranking: $\#231$ in DeFi, $\#639$ in General). As a lending protocol, {\em Aave} allows users to supply assets and earn passive income.

\begin{table}[]
\centering
\resizebox{\columnwidth}{!}{%
\begin{tabular}{|c|c|c|c|c|}
\hline
\textbf{File} & \textbf{Source\_Contract} & \textbf{Source\_Function} & \textbf{Target\_Contract} \\ \hline
WETH9Mock.sol & WETH9Mock & mint & WETH9Mock \\ \hline
WETH9Mock.sol & WETH9Mock & mint & None \\ \hline
MockBadTransferStrategy.sol & MockBadTransferStrategy & constructor & MockBadTransferStrategy \\ \hline
MockBadTransferStrategy.sol & MockBadTransferStrategy & performTransfer & MockBadTransferStrategy \\ \hline
... & ... & ... & ... \\ \hline
\end{tabular}%
}
\caption{Example of dataset returned by the tool for the dApp {\em Aave} (Ethereum - DeFi).}
\label{tab:output parser}
\end{table}

\begin{figure}[]
    \centering \includegraphics[width=0.7\textwidth]{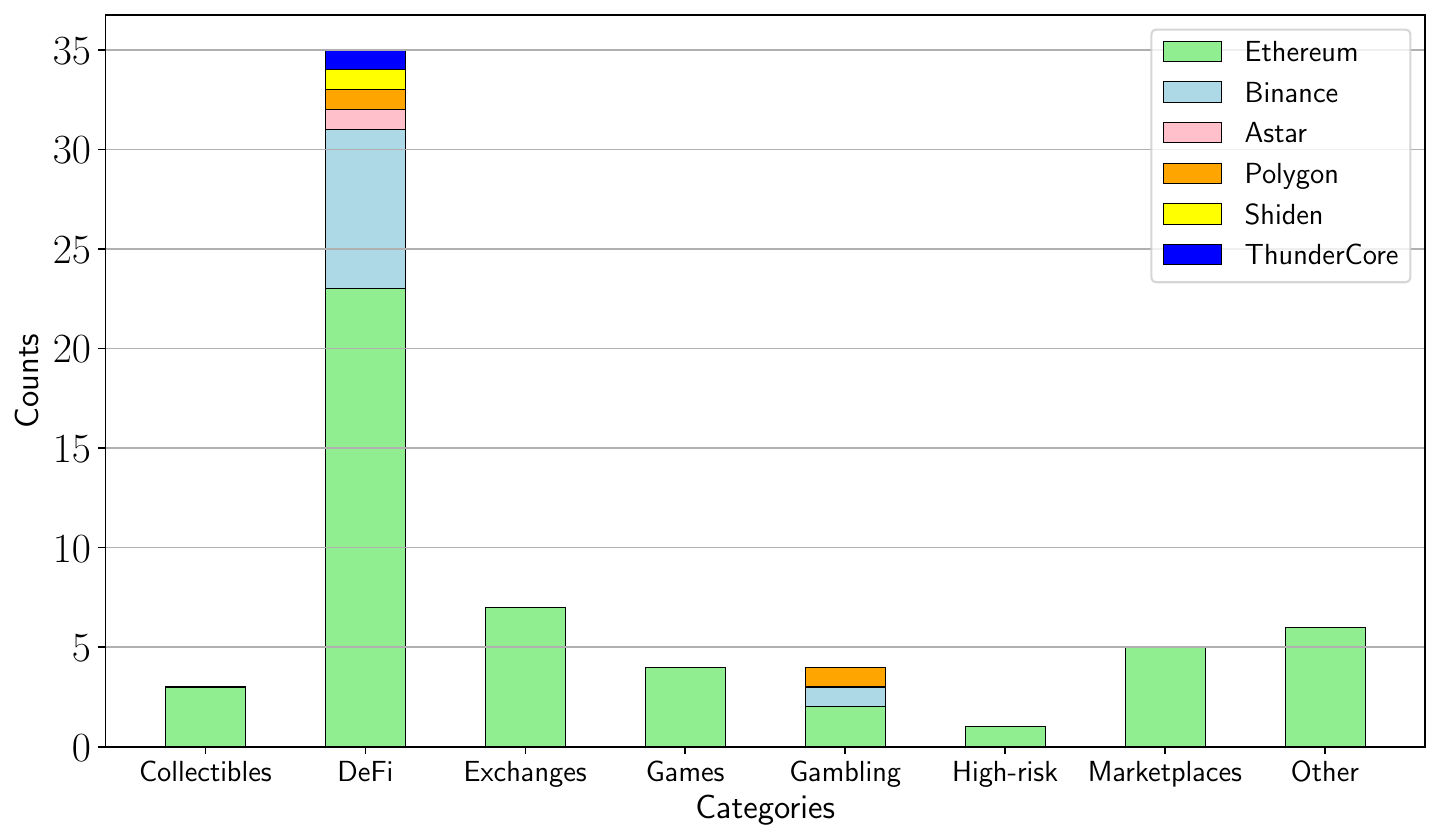}
    \caption{Composition of the dataset in terms of dApp's category/use-case and compatible blockchain.} 
    \label{fig:dataset composition}
\end{figure}

\begin{figure}[]
    \centering \includegraphics[width=0.7\textwidth]{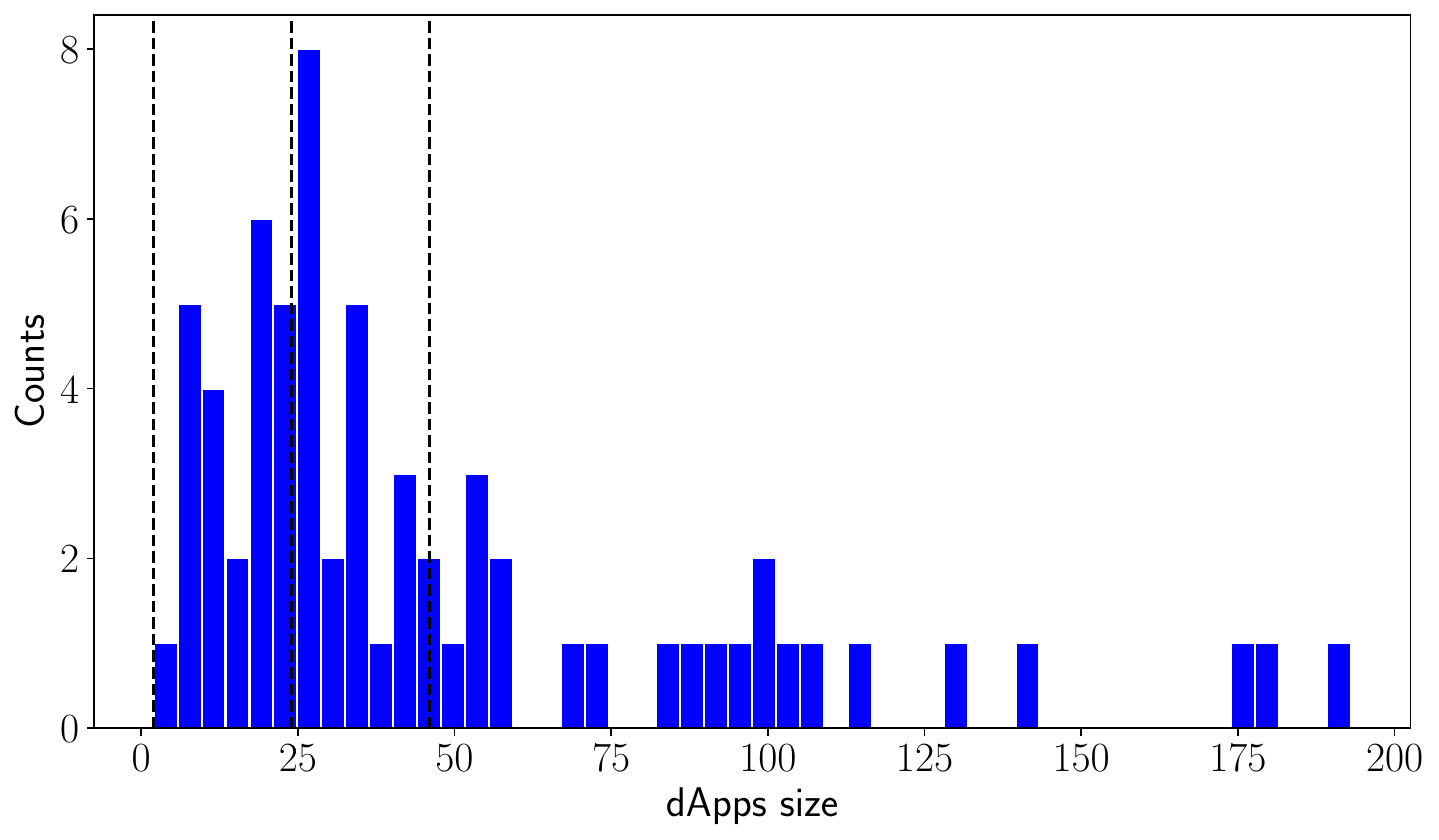}
    \caption{Distribution of dApp sizes. The number of {\em Source Contracts} is a proxy for the size. Our dataset is composed of $20$ Small dApps, $22$ Medium dApps, and $22$ Large dApps.}
    \label{fig:size dapp}
\end{figure}

\begin{figure}[]
    \centering
    \includegraphics[width=0.7 \textwidth]{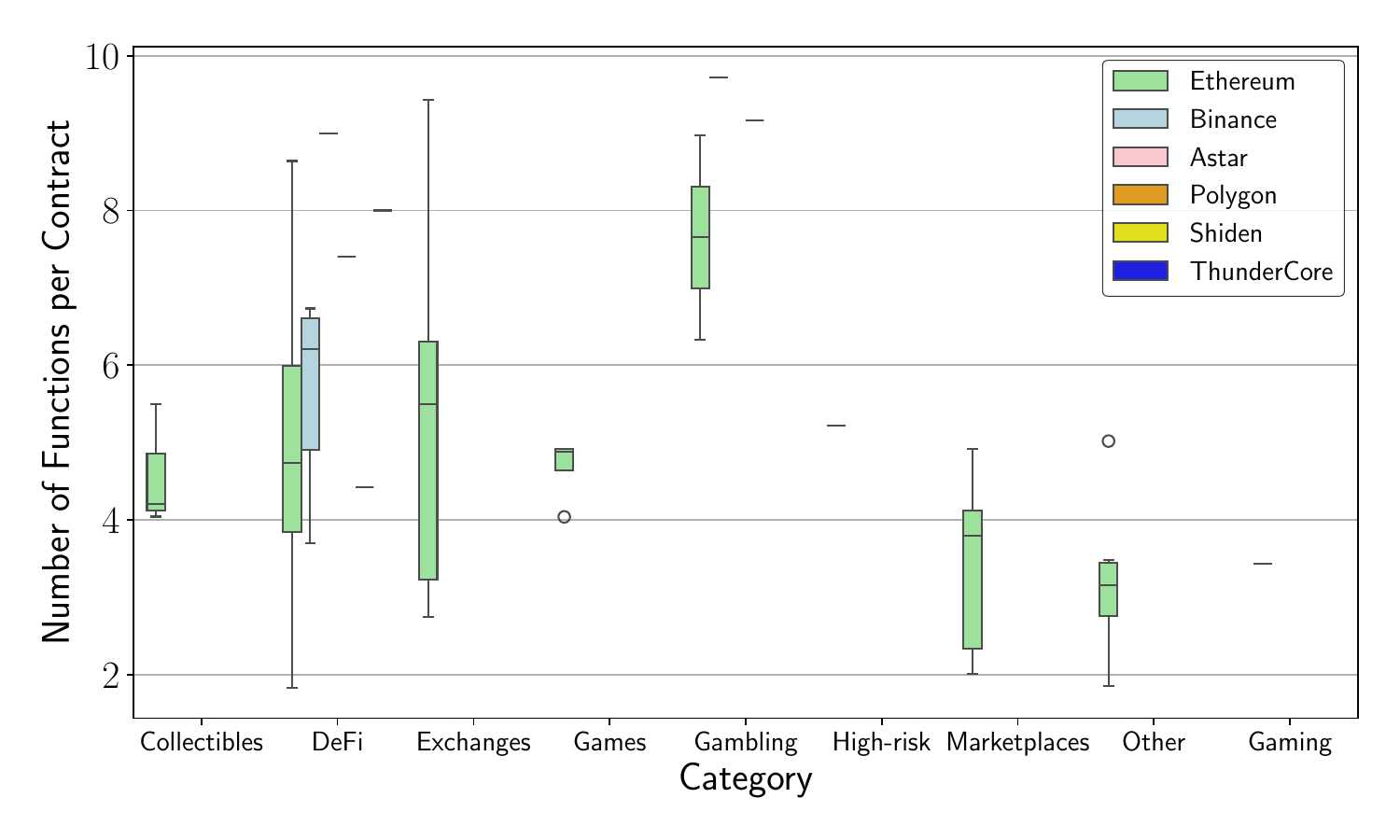}
    \caption{Box plot of the number of functions per contract. The ends of the box represent the first and third quartiles, the median (second quartile) is marked by a line inside the box, and the end of the whiskers represent the minimum and the maximum.}
    \label{fig:box plot}
\end{figure}

Overall, our dataset consists of $51$ dApps Ethereum-based, and $15$ dApps deployed on other blockchains (see Fig. \ref{fig:dataset composition} and Appendix \ref{app:list of dapps}). The majority of dApps are Ethereum-based, due to the significant expansion of the Ethereum ecosystems in recent years \cite{harvey2021defi}. These applications are related to multiple sectors, such as: \\
\textbf{DeFi} Applications in this category handle various aspects of financial services: Insurance, Investments, Lending and Borrowing, Payments, Token Swap, and Trading and Prediction Market. Each of these sub-categories brings a unique set of functionalities, all aiming to disrupt traditional financial systems by introducing automation, transparency, and efficiency through blockchain and smart contract technology. \\
\textbf{Art and Collectibles} This category of dApps focuses on digital ownership and artistic creation. Tokenization, based on so-called Non-Fungible Tokens (NFTs), plays a critical role in use cases related to establishing ownership and provenance. \\
\textbf{Gaming} These dApps offer interactive entertainment and virtual exploration, and they are generally divided into Competition and Digital World sub-groups. They allow buying and trading digital assets that can enhance gameplay, and they operate in environments that simulate various landscapes. \\
\textbf{Technology} This category contains dApps that aim to revolutionize developers' tools and integrate blockchain into existing technology platforms. They support open-source development initiatives and facilitate the decentralization of various technological services. \\
\textbf{Gambling} Gambling dApps comprise platforms allowing users to bet their money on gambling and high-risk games. They range from decentralized casinos to prediction markets. \\
\textbf{Staking} Staking dApps are  decentralized applications that allow users to lock their cryptocurrencies to support network operations, often in exchange for rewards or other benefits. It offers a way for users to potentially earn returns on their crypto holdings by participating in network security or governance. 
\\
We decided to concentrate on DeFi-related dApps as, in 2021, DeFi protocols emerged as the predominant targets of cryptocurrency hackers, and this pattern further intensified in 2022 \cite{crypto_crime_report_2023}.
\\
The number of Source Contracts that compose a dApp is taken as a proxy for its size, following \cite{ibba2018preliminary}: dApps are categorized into Small (3 to 23 contracts), Medium (24 to 45 contracts), and Large (46 to 193 contracts). The categorization of dApps into Small, Medium, and Large groups -- based on the number of contracts they are formed of -- is a heuristic approach driven by the characteristics of our dataset. The specific ranges (3 to 23, 24 to 45, 46 to 193) for these categories were selected to create a balanced sub-division that allows meaningful comparison and analysis across groups. We, therefore, have $20$ Small dApps, $22$ Medium dApps, $22$ Large dApps (see Fig. \ref{fig:size dapp}). 
This metric offers a quantitative measure of a dApp’s complexity and potentially its functional diversity. The smallest dApp is {\em 1inch Network} (Ethereum - DeFi) with $6$ functions and $2$ contracts, while the largest is {\em Balancer} (Ethereum - Exchanges) with $531$ functions and $193$ contracts.
\\
Functions are the fundamental building blocks of contracts and, therefore, dApps. In Fig. \ref{fig:box plot}, we analyse the number of functions in each dApp, adjusted for their respective sizes. On average, the number of functions is $5.09$ times the number of contracts within the same dApp, with a standard deviation of $1.91$. The minimum value is $1.83$ for the Ethereum - DeFi category, while the maximum value is $9.43$ for the Ethereum - Exchanges sector. The DeFi and Exchanges categories show the highest dispersion, with values exhibiting significant deviations from the median. The Gambling category is, instead, characterised by a larger number of functions on average. The presence of numerous functions suggests the reliance on multiple separate scripts to accomplish different tasks. This technique of splitting larger tasks into multiple sub-functions suggests a reduction in the responsibility of single contracts. If all tasks were concentrated in few functions, the likelihood of the dApp ceasing to function in the event of technical malfunctions would be considerably higher. Instead, by distributing tasks across a greater number of functions, the risk of a technical malfunction affecting the entire dApp is minimized. This result is not surprising, as a similar approach is observed in `standard' (i.e., non blockchain-related) software engineering.
In software engineering, the principal mechanism employed for designing object-oriented software is the class. The allocation of responsibilities and collaborations among classes can take various forms. In a delegated control style, a well-defined set of responsibilities is spread across multiple classes. These classes assume distinct roles and occupy recognized positions within the application architecture. Object-oriented design experts suggest that a delegated control style is more comprehensible and adaptable than a centralized control style \cite{arisholm2004evaluating}. This approach shares similarities with our findings in the context of dApps, where a distributed approach to functions mitigates the impact of potential technical failures on the entire system.
\\

\subsection{Building contract and function networks}
\begin{figure}[]
    \centering
    \includegraphics[width=0.6\textwidth]{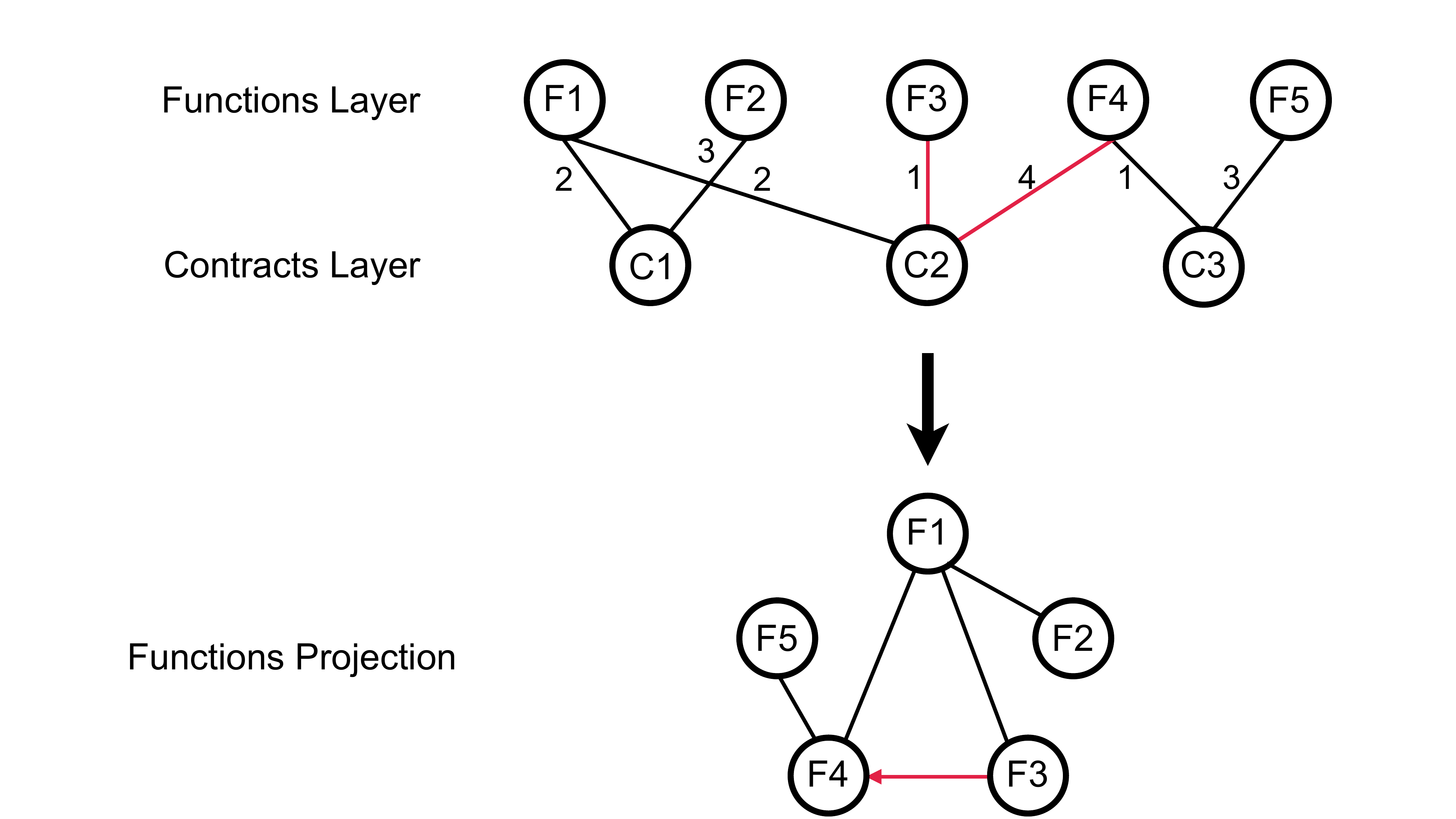}
    \caption{Illustration of the bipartite network {\texttt Functions - Contracts}, and its projection on the {\texttt Function} layer. For example, the edge weight from node $F3$ to $F4$ is computed as follows: it is equal to $1$ (the number of possible directions when starting from node $3$, weighted by the probability of going through that link) multiplied by the probability of reaching $F4$, divided by the total number of nodes through which the information can flow (weighted by their respective probabilities): $1 \cdot \frac{4}{2+1+4} = 0.57$.}
    \label{fig:proiezione}
\end{figure}

As shown in Tab. \ref{tab:output parser}, the dataset reveals interactions between the calling function and the contracts, which the call originates from and terminates into. We are indeed interested in conducting a more fine grained analysis, as our objective is to understand interactions {\em within} contracts and interactions {\em within} functions (considering functions with the same Source Contract as identical).
To construct the contracts network for each dApp, we use information on the Source and Target Contract of each function call. We build the network's adjacency matrix, where the rows denote the sources and the columns denote the targets (Tab. \ref{tab:contract matrix}). 
\begin{table}[]
\centering
\resizebox{\columnwidth}{!}{%
\begin{tabular}{|c|c|c|c|c|c|}
\hline
 & \textbf{ACLManager} & \textbf{AToken} & \textbf{ATokenHarness} & \textbf{AaveEcosystemReserveController} & ... \\ \hline
\textbf{ACLManager} & 1 & 0 & 0 & 0 &  \\ \hline
\textbf{AToken} & 0 & 9 & 0 & 0 &  \\ \hline
\textbf{ATokenHarness} & 0 & 0 & 1 & 0 &  \\ \hline
\textbf{AaveEcosystemReserveController} & 0 & 0 & 0 & 5 &  \\ \hline
... &  &  &  &  &  \\ \hline
\end{tabular}%
}
\caption{Example of adjacency matrix for {\em Aave}'s (Ethereum - DeFi) contracts network.}
\label{tab:contract matrix}
\end{table}
The matrix element at position \textit{(i,j)} may assume a value of $0$, if the function belonging to the Source Contract \textit{i} does not call Target Contract \textit{j}, or it may assume a value of $n \in \mathbb{N}$, if the function in Source Contract \textit{i} calls Target Contract \textit{j} $n$ times. We, thus, obtain for each dApp a weighted directed network of contracts interactions, resulting in $66$ contracts networks. \\
To infer the network of connections among functions, further steps are necessary. We use the information regarding the Source Function and the Target Contract: analysing the relationship between them is crucial to determine the system's robustness. Indeed, vulnerabilities in function-contract calls have been exploited in hacking attacks aimed for instance at stealing funds from cryptocurrency wallets and applications \cite{destefanis2018smart,sayeedsmart}. 
As previously done, we build the bi-adjacency matrix, where the rows denote the Source Functions and the columns denote the Target Contracts. The matrix element at position \textit{(i,j)} may assume a value of $0$, if function \textit{i} does not call target contract \textit{j}, or a value of $n$, if function \textit{i} calls target contract \textit{j} $n$ times. Given the possibility of multiple calls from the same function to the same contract, we obtain a weighted bipartite graph. The two layers are Functions (layer $F$) and Contracts (layer $C$) as schematically depicted in Fig. \ref{fig:proiezione}, top panel.
Since our interest lies in the relationships {\em within} the functions layer, we project the information onto the single layer $F$. The one-mode projection onto layer $F$ results in a network consisting exclusively of $F$ nodes, and it is a procedure extensively used in graph theory.
Determining how to weight the edges in this network is a critical aspect of the one-mode projection. We adopt a methodology similar to the one introduced by Tao Zhou et al. in \cite{zhou2007bipartite}. In order to assess if the contract $c \in C$ called by function $f_1 \in F$ is more likely to be called also by function $f_2 \in F$ we have to perform a contraction of the bi-adjacency matrix $M$ over the contract dimension, i.e., the set $C$ of contracts, and take the element $(f_1, f_2)$. This method is called {\em probabilistic spreading approach}. Let us consider one bit of information on a generic function $f_1 \in F$. We aim to describe how this information can spread to contracts in $C$, then back to $F$. Firstly, the information moves to the contracts layer according to the connection patterns of $M$. The probability that the information goes from $f_1$ to a given contract $c$ is
\begin{equation}
    \rho_{f_1 \rightarrow c} ^ {F \rightarrow C} = \frac{M_{f_1, c}}{\sum_{\tilde{c} \in C}M_{f_1, \tilde{c}}} \ , 
\end{equation}
where $\sum_{\tilde{c} \in C}M_{f_1, \tilde{c}}$ is the number of possible paths from $f_1$ to $C$, each weighted by the probability of going through a given path. Since the elements of $M = 0,1,\ldots, n$ we are not assuming equal transition probabilities, introducing a bias in the process. Secondly, the information that reached the contracts layer jumps back to the functions one, following again the connection patterns of $M$. The transition probability from $c$ to a given function $f_2$ in layer $F$ is:
\begin{equation}
     \rho_{c \rightarrow f_2} ^ {C \rightarrow F} = \frac{M_{f_2, c}}{\sum_{\tilde{f} \in F}M_{\tilde{f}, c}}   \ ,
\end{equation}
where $\sum_{\tilde{f} \in F}M_{\tilde{f}, c}$ is the number of possible weighted paths from $c$ in layer $C$ to layer $F$. Finally combining these steps, the probability that the bit of information jumps from function $f_1 \in F$ to function $f_2 \in F$, via all possible connected contracts $c$, is
\begin{equation}
    \rho_{f_1 \rightarrow f_2} ^ {F \rightarrow F} =     {\sum_{c \in C} \rho_{f_1 \rightarrow c} ^ {F \rightarrow C} \rho_{c \rightarrow f_2} ^ {C \rightarrow F} } = \frac{M_{f_1, c}}{\sum_{\tilde{c} \in C}M_{f_1, \tilde{c}}} \frac{M_{f_2, c}}{\sum_{\tilde{f} \in F}M_{\tilde{f}, c}}  \ .
    \label{matrici da filtrare}
\end{equation}
Eq. \eqref{matrici da filtrare} defines a monopartite network of $F$ nodes, which can be interpreted as the flow of information {\em within} functions in $F$. We can interpret the connections of this network as conditional probabilities $P(f_2 | f_1) = {\sum_{c \in C}} P(f_2|c) P(c|f_1)$. \\
We thus obtain for each dApp a weighted directed monopartite network of functions interactions, resulting in $66$ functions networks. In Tab. \ref{tab:function matrix} an example of the bi-adjacency matrix for {\em Aave}'s functions network.
\begin{table}[]
\centering
\resizebox{\columnwidth}{!}{%
\begin{tabular}{|c|c|c|c|c|c|}
\hline
 & \textbf{\_approveDelegation} & \textbf{\_approve} & \textbf{\_approve\_Incentivized} & \textbf{\_burnScaled} & ... \\ \hline
\textbf{\_approveDelegation} & 0.009 & 0.009 & 0.009 & 0.03 &  \\ \hline
\textbf{\_approve} & 0.009 & 0 & 0.06 & 0.03 &  \\ \hline
\textbf{\_approve\_Incentivized} & 0 & 0 & 0 & 0.11 &  \\ \hline
\textbf{\_burnScaled} & 0.004 & 0.004 & 0.004 & 0.018 &  \\ \hline
... &  &  &  &  &  \\ \hline
\end{tabular}%
}
\caption{Example of bi-adjacency matrix for {\em Aave}'s (Ethereum - DeFi) functions network.}
\label{tab:function matrix}
\end{table}

To assess the statistical significance of the elements of the matrices defined in \eqref{matrici da filtrare}, we resort to a null model. We use the disparity filter\footnote{\url{https://github.com/DerwenAI/disparity_filter}}, a filtering method that extract the relevant connection backbone in complex networks, preserving the edges that represent statistically significant deviations with respect to a null model for the local assignment of weights to edges \cite{serrano2009extracting}. An important aspect of this method is that it does not affect small-scale interactions and operates at all scales defined by the weight distribution. In this context, the information on weights is significant as they directly correlate with the frequency of functions calling a contract, influencing the associated gas fee expenses. We adopt a filtering method that retains edges that represent statistically significant deviations when compared to a null model of local weight assignment. It filters out connections characterized by substantial disorder, while preserving structural properties and hierarchies. Our findings from the network analysis indeed reflect intrinsic characteristics of the systems we are examining, rather than being a mere consequence of the chosen filtering method. 
In Sec. \ref{sec:Results}, we report the results of the analysis conducted on the filtered weighted functions and contracts networks.
 
\section{Results} \label{sec:Results}

We analyse $51$ Ethereum-based dApps and $15$ dApps from other blockchains (Binance, Optimism, Polygon, Astar, Shiden, and ThunderCore), spanning various categories, and of varying sizes. 
\\

\begin{figure}
    \centering
    \includegraphics[width=0.6\textwidth]{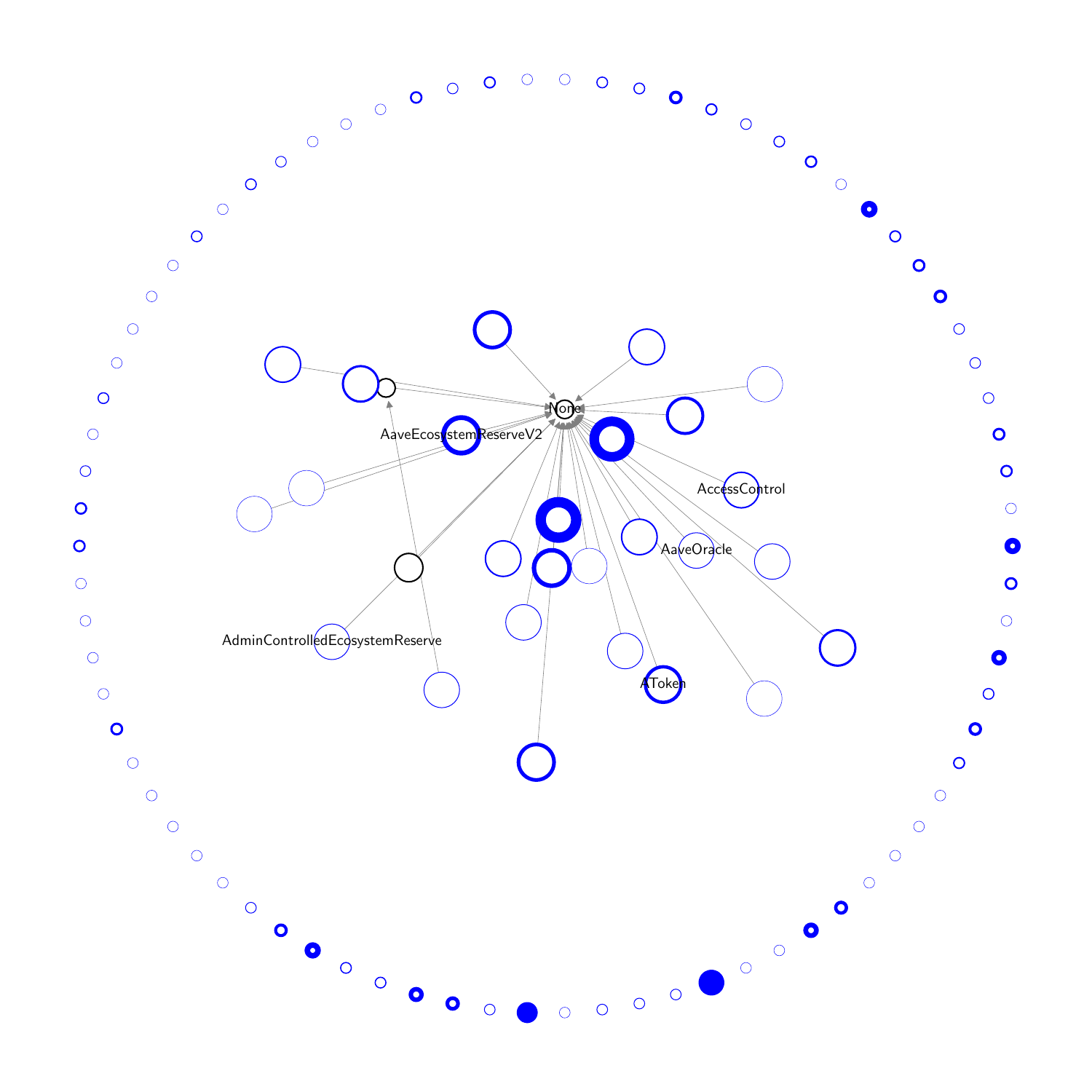}
    \caption{Network of contracts' interactions for {\em Aave}. In blue, the nodes with self-loops. The width of the border of each blue node corresponds to the weight of its self-loop link. The nodes positioned on the external circle do not interact with other contracts.}
    \label{fig:aave contratti}
\end{figure}

\begin{figure}[]
    \centering
    \includegraphics[width=0.9\textwidth]{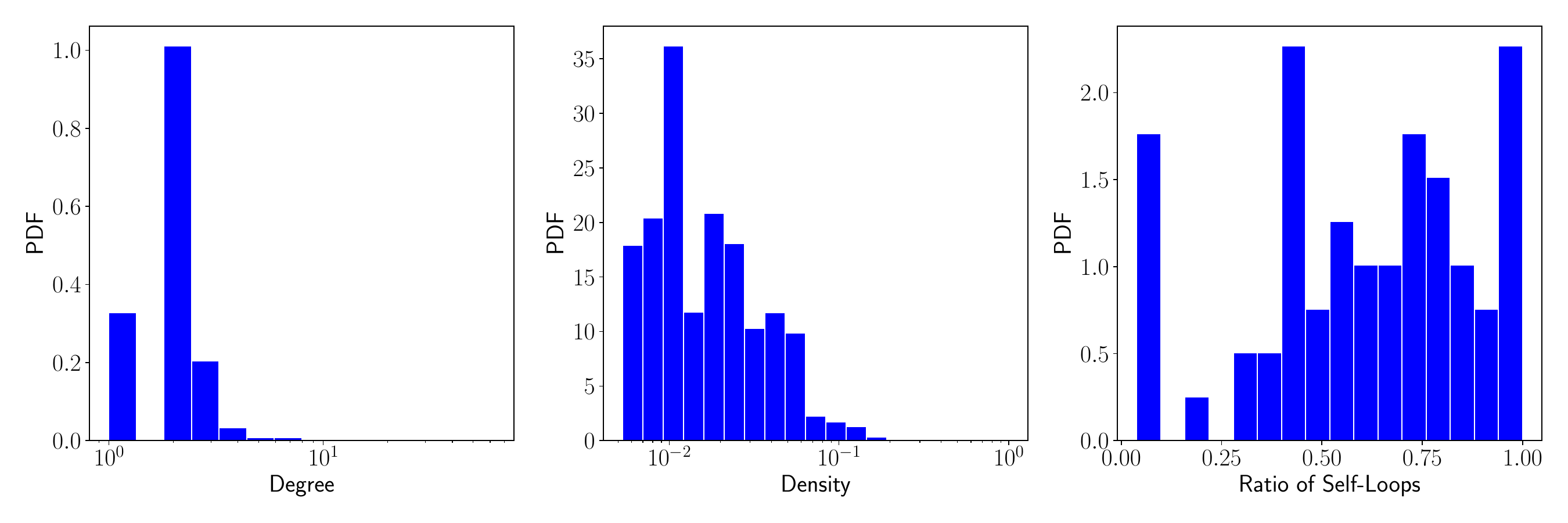}
    \caption{Left Panel: Probability Density Function (PDF) of the degree distribution for the $66$ contracts networks. Central Panel: PDF of the density of the networks. Right Panel: For each network, the ratio of nodes having a self-loop as their sole connection to the total number of nodes is presented.}
    \label{fig:PDF}
\end{figure}

\textbf{Contracts Networks} 
We generate a total of $66$ weighted directed networks illustrating contracts' interactions, with each network representing a dApp.
In these networks the nodes represent a contract, and the width of the links reflects the strength of interactions from the source to the target contract (how many times it is called), while the node sizes are scaled based on the number of target contracts that a given contract calls.
In Fig. \ref{fig:aave contratti} the network of contracts' interaction for {\em Aave} is presented.  The names of the contracts with the highest betweenness centrality are listed. For instance, {\em None} is a {\em Context} contract, a dependency used to return the contest of transaction sender and data.
In Fig. \ref{fig:contracts} (Appendix \ref{app:contracts networks}) several other examples of contracts networks are presented. We decide to present the networks of dApps that, at present, exhibit notable balances – defined as the total value of current assets held in the dApp's smart contracts. To ensure a representative sample, we include examples from various categories and blockchain platforms.
Fig. \ref{fig:PDF} shows the distribution of networks' degrees and density, and the prevalence of self-loops as the only connections. The networks exhibit a low degree, with the majority falling within the range of $2$ to $3$: the nodes have few connections. The Left and Central panels reveal consistent characteristics within the contracts networks of dApps, regardless of their size or category, specifically their sparsity.  The density of a graph is a measure of how many potential edges are present in the graph compared to the total number of possible edges in a complete graph of the same size. The networks display a low density, with the exception of {\em 1inch Network}, which has a density of $1.0$. However, as said in Sec. \ref{sec:Methodology}, {\em 1inch Network} has only $2$ contracts, making this result quite trivial. In the Right panel, it becomes apparent that the majority of links consists of self-loops. A total of $54$ networks have a minimum of 40\% of nodes with self-loops, and $33$ networks (more than half of the dataset) have at least 60\% of their nodes connected solely through self-loops.  A Louvain modularity analysis on the undirected version of these graphs produces an average modularity coefficient of $0.8$ across all networks. In conclusion, the networks exhibit evident sparsity, with the majority of connections being self-loops, and community structure is highly significant, resulting in a lot of distinct components. \\
In a software engineering framework, it means that the dApps are designed with a high level of independence and minimal inter-contract dependencies. This choice may be a deliberate strategy to improve security and reduce the risk of chain failure, given the immutability of contracts when deployed. The presence of self-loops indicates that most contracts are self-sufficient, executing functions and maintaining a state without the need for external calls or interactions. Instead, the presence of few communities indicates sets of contracts grouped by functionality or purpose, facilitating maintainability and potential scalability. 

\vspace{2pt}

\begin{figure}[H]
    \begin{minipage}{0.5\textwidth}
        \centering \includegraphics[width=\textwidth]{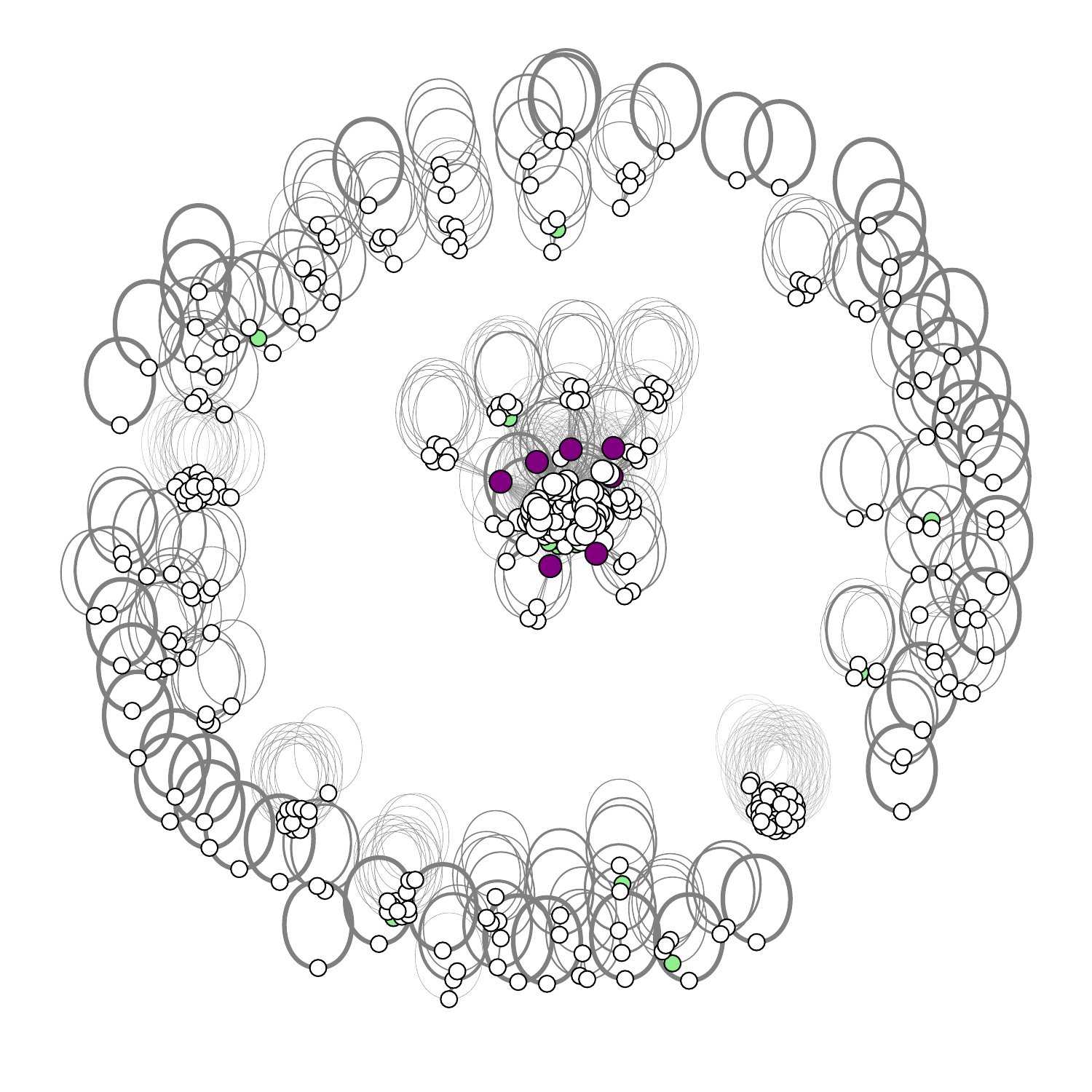} \subcaption{Network for functions interactions pre-filter}
    \end{minipage}%
    \begin{minipage}{0.52\textwidth}
        \centering \includegraphics[width=\textwidth]{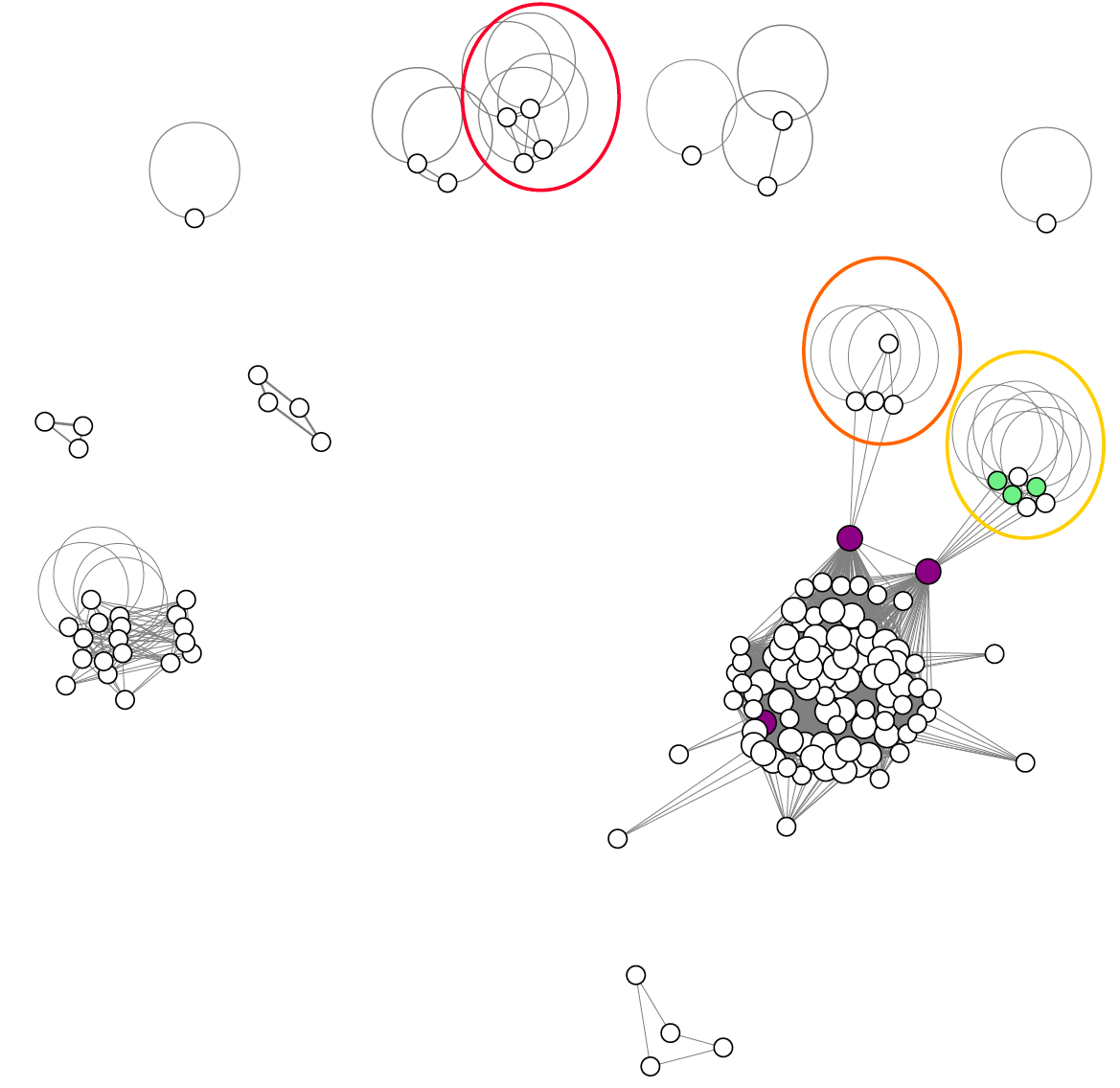} \subcaption{Network for functions interactions post-filter}
    \end{minipage}
    \caption{{\em Aave} Functions Network (Ethereum - DeFi). In purple the nodes with highest betweenness (i.e. {\em cancel}, {\em withdrawFromStream}, {\em setClaimer}), in green the nodes with highest clustering coefficient (i.e. {\em executeDelegateCall}, {\em updateDelay}, {\em updateGracePeriod}). The purple functions are responsible for authorizing an address to withdraw tokens on behalf of another specific address, clearing a queue of actions for execution, or making a token withdrawal from a Stream. The green functions are responsible for updating the time (setting the value to the next moment when the execution of a set of actions concludes) and adjusting the delay between queuing a set of actions and their execution.
    } \label{fig:aave funzioni}
\end{figure}

\textbf{Functions Networks}
We generate a total of $66$ weighted directed networks illustrating functions' interactions, with each network corresponding to a specific dApp. On average, the ratio between post-filter and pre-filter nodes stands at $65\%$, with a standard deviation of $14\%$. The minimum ratio of $32\%$ is observed in the case of  {\em SWAPP Protocol} (Ethereum - DeFi), which features $214$ nodes pre-filter (i.e. functions) reduced to $71$ nodes post-filter. In contrast, {\em Plexus} (Ethereum - Exchanges) shows the maximum ratio of $97\%$, having initially $66$ nodes, which are reduced to $64$ post-filter.
The networks are visualized using the  {\em spring} layout; the nodes represent functions, and the width of the links reflects the strength of interactions between functions. The node sizes are adjusted according to the number of target contracts called by each function, and nodes are displayed in a purple shade if they rank among the nodes with the highest betweenness, while a green color is assigned to the nodes with the highest clustering coefficient. In Fig. \ref{fig:aave funzioni} the network of functions interactions for {\em Aave} is presented. We observe a set of smaller components that represent the secondary functionalities of the dApp. The characteristic of the contract network having a high number of self-loops translates into this network as all minor components consisting of functions defined within the same contract. For instance, in the red component, there are only functions defined within the {\em PolygonBridgeExecutor} contract. In the orange community functions are defined within {\em AaveEcosystemReserveV2}, and in the yellow community they belong to {\em BridgeExecutorBase}. In contrast, the core of the largest component comprises functions defined in multiple contracts, representing the main functionality of the dApp. For instance, within it, we find functions such as {\em setReserveInterestRate}, {\em setPoolImpl}, and {\em setLiquidationProtocolFee}, of fundamental importance for the functioning of the dApp.
In Fig. {\ref{fig:balancer funzioni}, \ref{fig:venus funzioni}, \ref{fig:reality cards funzioni}} (see Appendix \ref{app:functions networks}) several other examples of these network visualizations are presented. The structure is consistent across all of them: there is always a largest component for the main functionality and a series of minor ones.
\subsection{Characteristics of networks of functions in dApps }
\begin{figure}[htb!]
    \centering
    \includegraphics[width=0.9\textwidth]{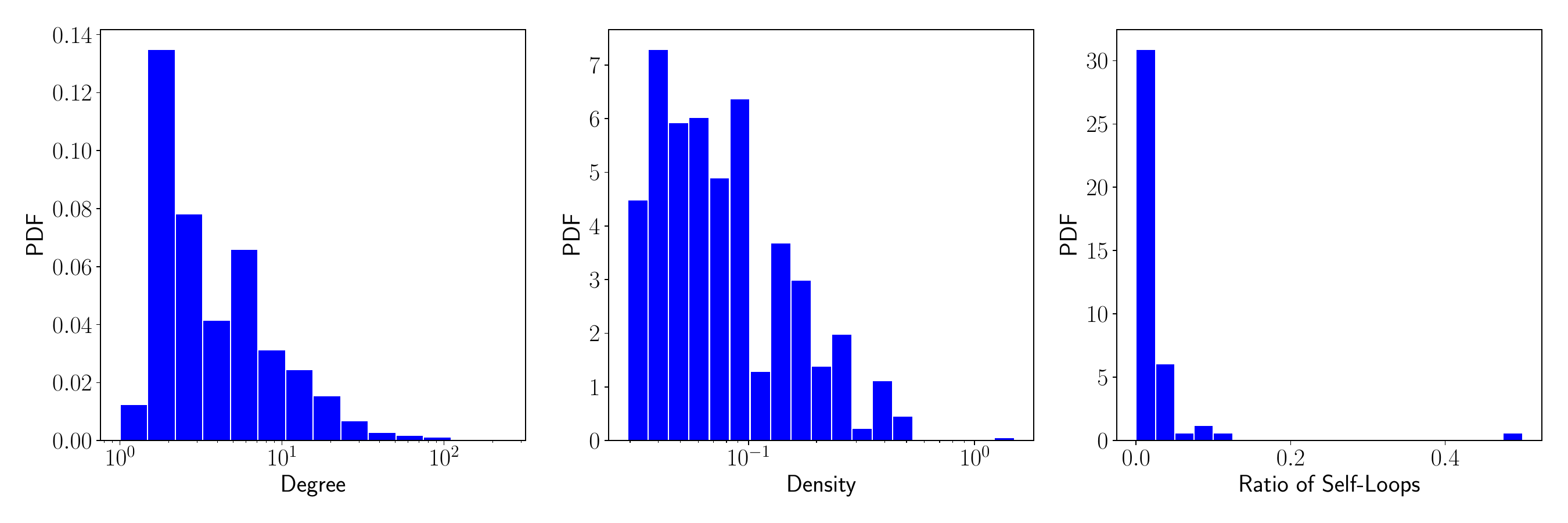}
    \caption{Left: PDF of the degree distribution for the $66$ functions networks. Central: PDF of the density  distribution. Right: For each network, the ratio of nodes having a self-loop as their sole connection to the total number of nodes.}
    \label{fig:PDF funzioni}
\end{figure}

In the case of functions networks, Fig. \ref{fig:PDF funzioni}, still reveals consistent characteristics across dApps, regardless of their size or category, but the scenario differs from what was observed previously with contracts. The degree of the functions networks is higher compared to the case of contracts networks, signifying a greater degree of connection.
 Similarly, the network density is higher, suggesting a lower sparsity in the graphs. In the case of {\em 1Inch Network}, the density is $1.6$, as self-loops are included in the total count of edges, which can result in densities exceeding $1$. 
Finally, the proportion of nodes having self-loops as their sole connection is significantly smaller compared to contracts networks: out of the $66$ networks, $65$ have a fraction of nodes connected solely by self-loops that accounts for $15\%$ of the total nodes. On average, an analysis of Louvain modularity within the networks gives a result of $0.69$, confirming reduced division into distinct components and highly connected nodes. This suggests greater complexity in interactions and a greater level of integration and task sharing among functions within the same dApp. 
\subsubsection*{Networks' Metrics} 
\begin{figure}[H]
    \centering
    \includegraphics[width=0.6\textwidth]{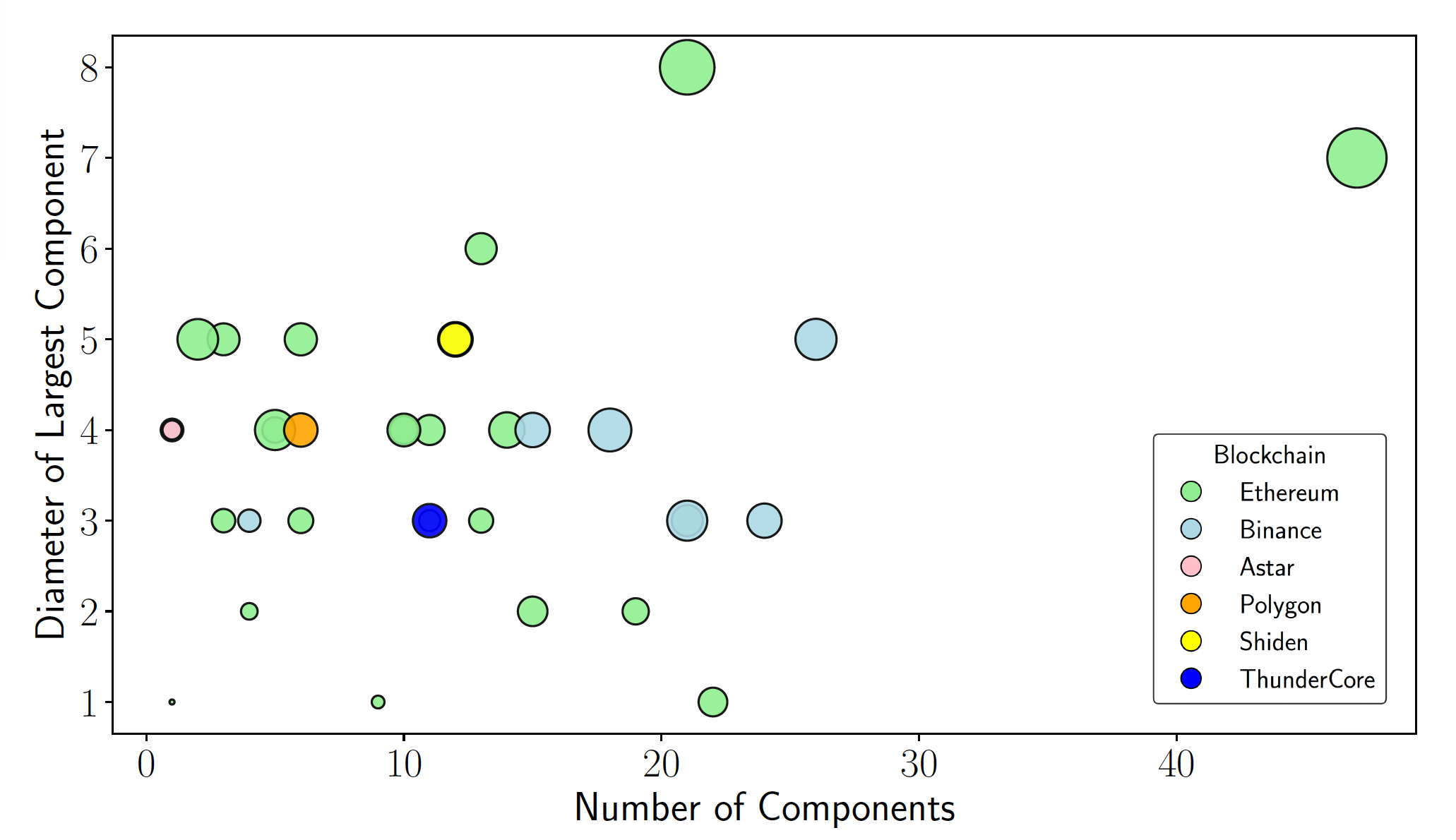}
    \includegraphics[width=0.6\textwidth]{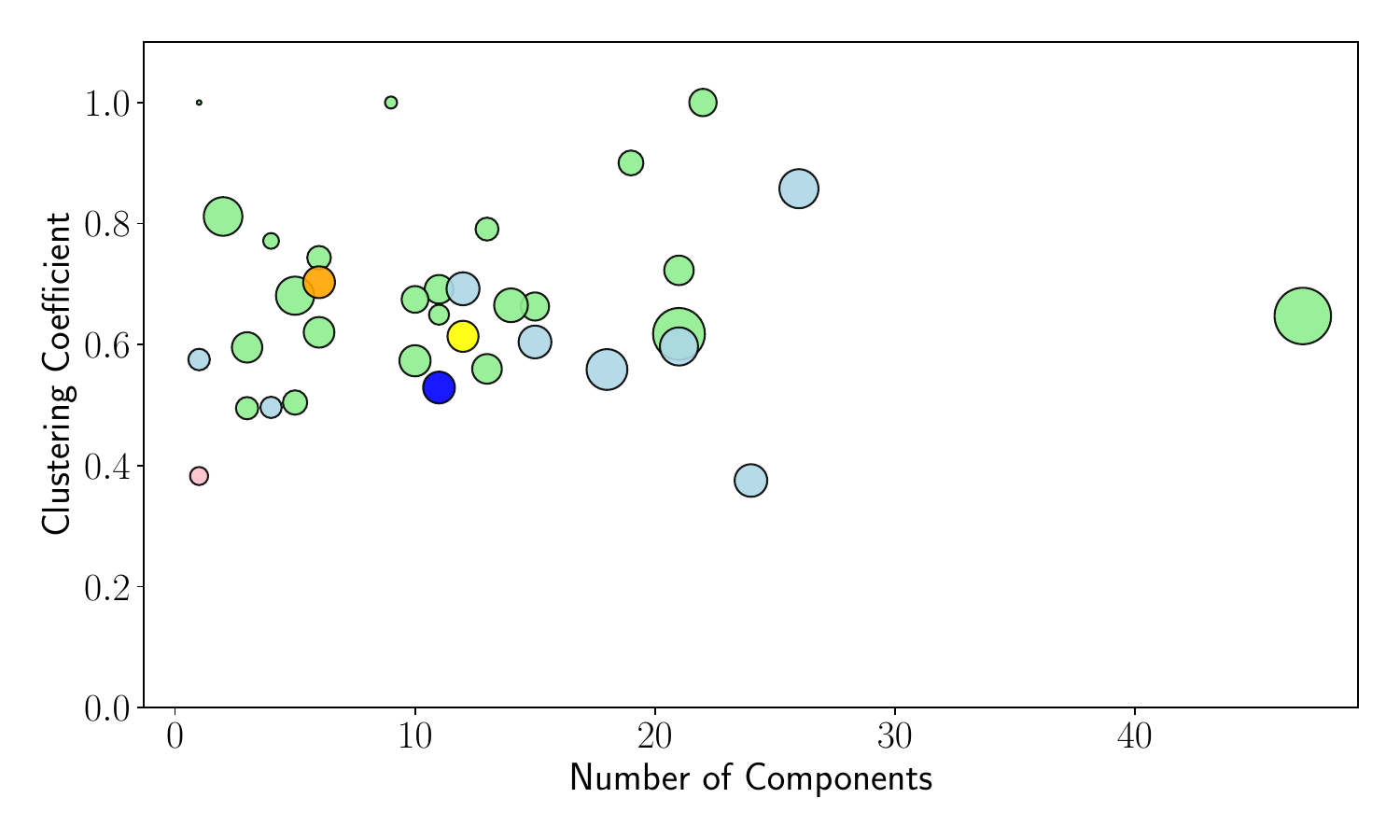}
    \caption{Top: Scatter plot of the diameter of the largest connected component as a function of the number of components. Bottom: Scatter plot of the global clustering coefficient in the largest components vs the number of components. Each dot represents a dApp, and the size of the dot is proportional to the number of functions in the specific dApp. This plot is restricted to dApps belonging to the DeFi category.}
    \label{fig:plot 1}
\end{figure}

\begin{figure}[H]
        \centering \includegraphics[width=0.45\textwidth]{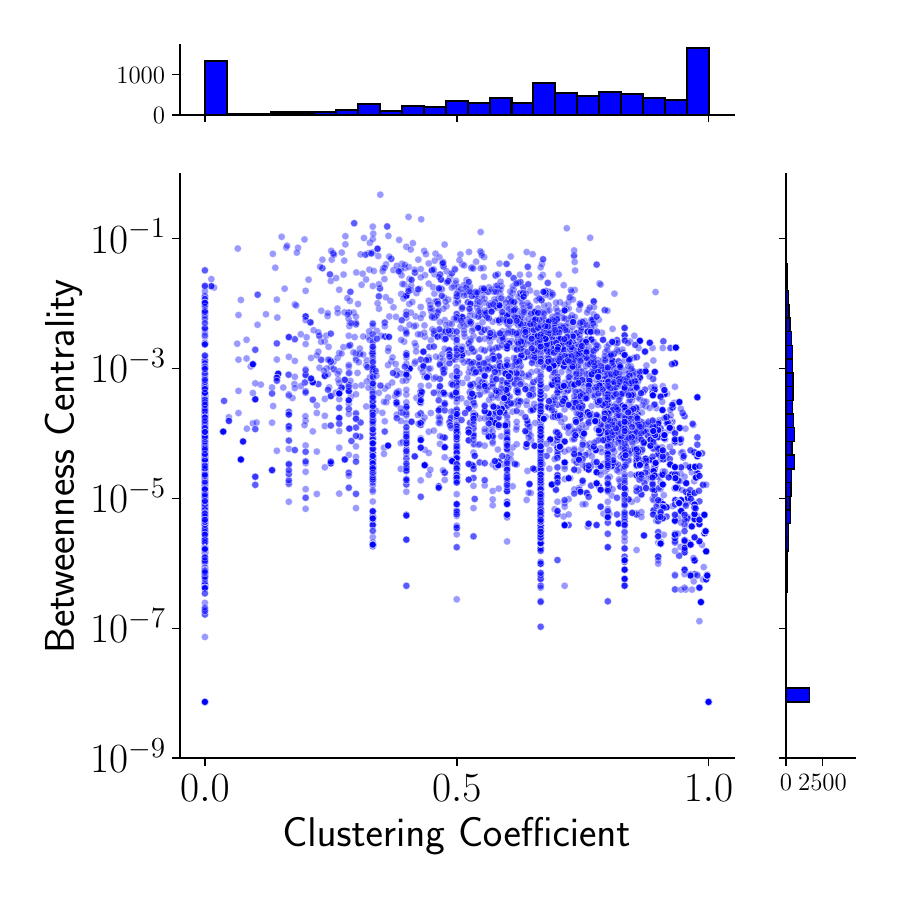} 
        \centering \includegraphics[width=0.45\textwidth]{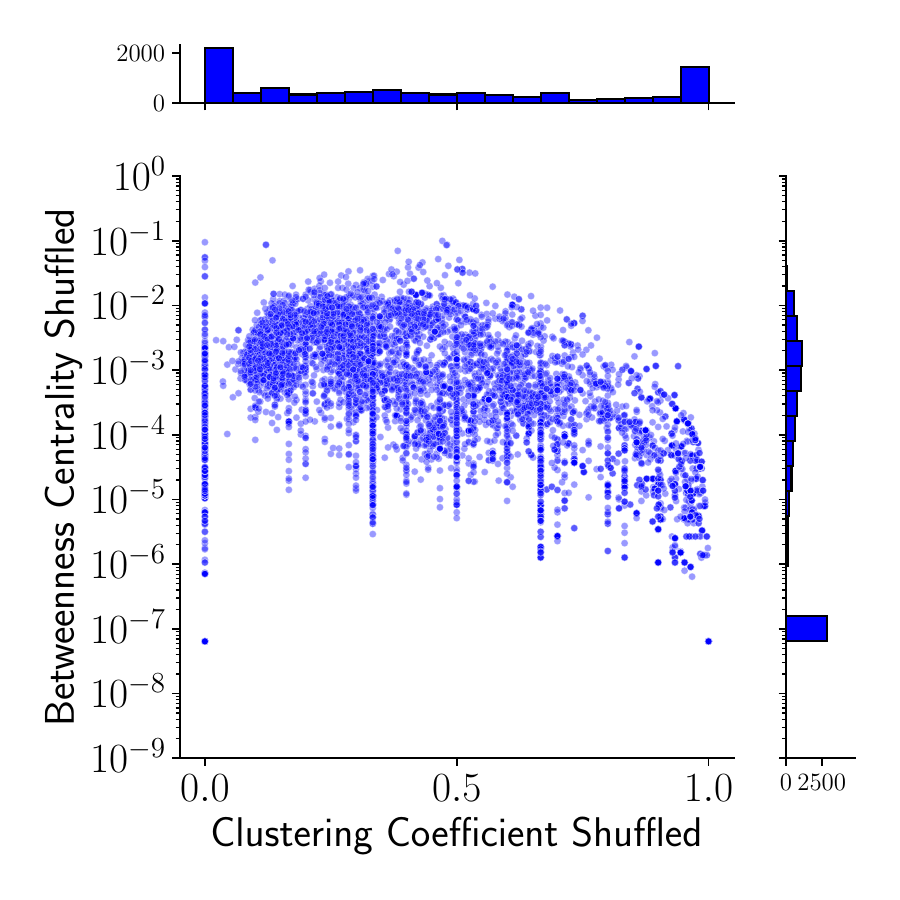} 
    \caption{Left: Relationship between clustering coefficient and betweenness in the original networks. Right: Relationship between clustering coefficient and betweenness in a random null model. } 
    \label{fig:betwenn vs clustering}
\end{figure}

\begin{figure}[H]
  \centering
    \includegraphics[width=0.45\textwidth]{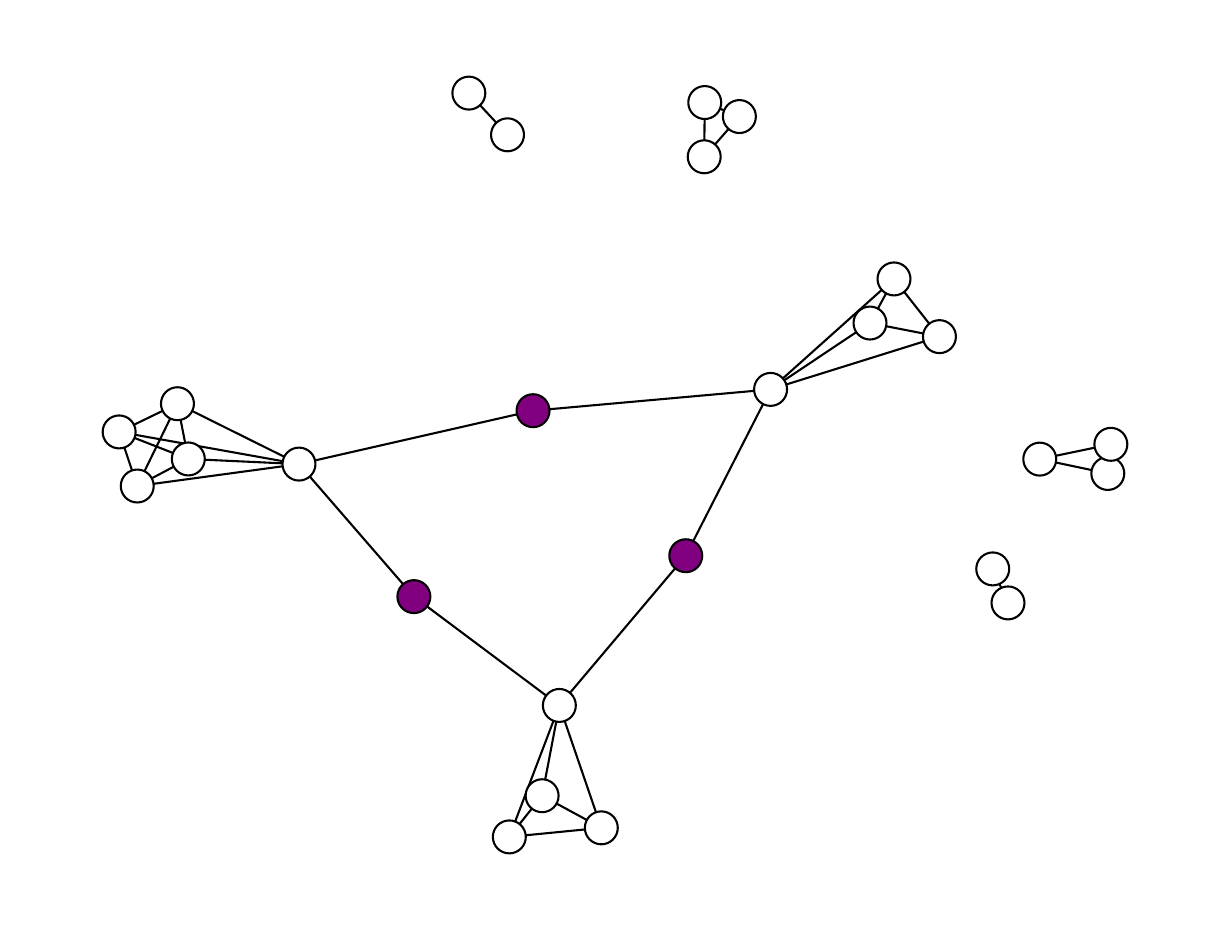}
       \includegraphics[width=0.45\textwidth]{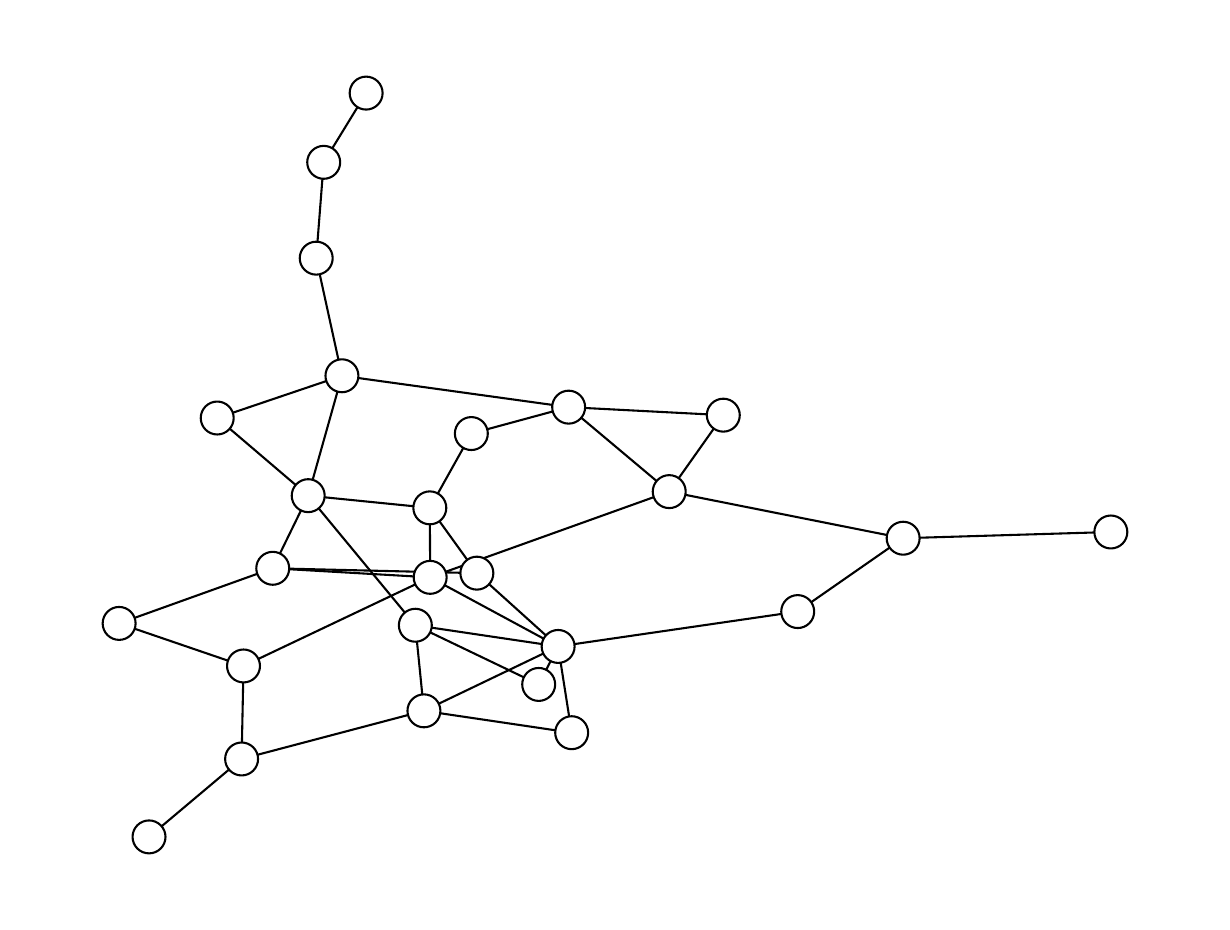}
    \caption{Left: Scheme of the typical real network structure of functions interactions. There is a largest component and few minor disconnected components on the side. The largest component is composed of different communities, with few bridges between them (the purple nodes). These sub-components are complete sub-graphs. Right: Randomised version of the function network with same number of nodes.}
    \label{fig:comparison}
\end{figure}

In Fig. ~\ref{fig:plot 1} - top panel, we show the relationship between the diameter of the largest connected component and the number of components in DeFi dApps is presented. Similar results and plots are for non DeFi dApps in Fig. ~\ref{fig:plot 2} in the Supplementary Material. Our emphasis is placed on the largest component as it represents the most crucial part of the network, housing the core functionalities of the dApp. Functions outside of this component perform less essential actions. The plots reveal consistent trends regardless of the dApp's category, affirming the presence of a common development pattern that is independent of the dApp's intended purpose. As previously mentioned, a distinct division into separate components is evident (with the number of components obviously increasing with the number of functions within the network). Functions within these dApps tend to form discrete groups with limited interactions between these groups, indicating a certain degree of compartmentalization in how functions are structured and interact within the application. Different sets of functions perform specific roles and maintain limited direct interactions with functions defined in other contracts. Additionally, as dApps increase in complexity by incorporating more functions (as indicated by the larger data points on the scatter plot), their internal network structure appears to become more intricate. The maximum number of connections within the largest component tends to grow with the number of functions in the dApp. Larger dApps may exhibit greater specialization, requiring a broader range of functions to manage specific tasks and fostering increased interaction and communication among these functions. 
The clustering coefficient is the fraction of all possible pairs of neighbors of node $i$ that are themselves linked in the graph. The clustering is like a local version of the betweenness, which is in turn a measure of centrality based on shortest paths. Betweenness and local clustering are, indeed, correlated \cite{newman2018networks}. If a vertex has a larger local clustering value, then the neighbors of the vertex can directly communicate with each other rather than going through the particular vertex. If the neighbors of a vertex do not need go through the vertex for shortest path communication, then it is more likely that the rest of the vertices in the network would not need to go through the vertex for shortest path communication. If a vertex has a smaller local coefficient, then the neighbors of the vertex are more likely to go through the vertex for shortest path communication between themselves (as there is more likely not a direct edge between the two neighbors, because of the low local coefficient for the vertex) \cite{burt2018structural}. Therefore, we expect an inverse relationship between these two measures in our networks. In Fig. \ref{fig:betwenn vs clustering}, we present an analysis of the relationship between betweenness and clustering coefficient of each node located within the largest component of each function network. Our goal is to discern whether there exists a distinctive characteristic specific to the dApps.
Given the consistent network patterns identified through our prior analyses, regardless of the dApp's category or the blockchain the dApp is deployed on, we choose to analyse all nodes across all function networks together. As expected, the plot exhibits a clear trend, reinforcing a notion of similarity and consistent structural patterns across the function networks. 
In the Left Panel, we illustrate the relationship between the two quantities in the original networks. In the Right Panel, we present the same relationship calculated for a null network model, constructed as configuration model retaining the same number of nodes and block structure while shuffling the links. For each network, we construct a random one, which retains the same modular structure as the original one, but connects nodes within each sub-component randomly. This null model preserves the modularity of the original networks, meaning that the compartmentalization of functions and the connections among functions performing similar tasks are maintained, but randomly reconnect functions designated for a particular task. This ensures that the results we obtain are not a mere consequence of the network's structural characteristics.
As expected, Fig. \ref{fig:betwenn vs clustering} shows that nodes with higher clustering coefficient tend to have lower betweenness centrality. This implies that nodes surrounded by highly connected neighbors, located in densely connected areas of the graph, are the same nodes with limited involvement in shortest paths between other nodes. As a result, they are not essential for the overall network connectivity nor efficient information transmission. Instead, nodes with high betweenness, serving as fundamental intermediaries for the flow of information or influence within the network, work as connectors between highly connected areas. However, when comparing our networks to the null model, we observe significant differences in the distribution of the clustering coefficient. In our networks, the mean clustering coefficient is higher, with the majority of data points falling within the range of $0.5$ to $1$. In contrast, the null model exhibits data points with lower clustering coefficients. Furthermore, the null model displays lower and less variable betweenness centrality values compared to the original real networks. A schematic representation of the structural differences between the real and randomised version of the networks can be seen in Fig. \ref{fig:comparison}. Therefore, our systems have characteristics that are independent of their modular structure. These characteristics include variable betweenness centrality with lower values and higher clustering coefficients, implying that our networks inherently consist of highly interconnected communities within each component, with only few nodes serving as bridges between them. 
In the function network examples in Figs. \ref{fig:balancer funzioni}, \ref{fig:venus funzioni}, \ref{fig:reality cards funzioni} in Appendix \ref{app:functions networks}, nodes with higher betweenness are marked in purple, while those with a higher clustering coefficient are marked in green. The purple nodes act as intermediaries between sub-components.
\subsubsection*{Small World Model}
\begin{figure}[H]
    \centering
    \includegraphics[width=0.7\textwidth]{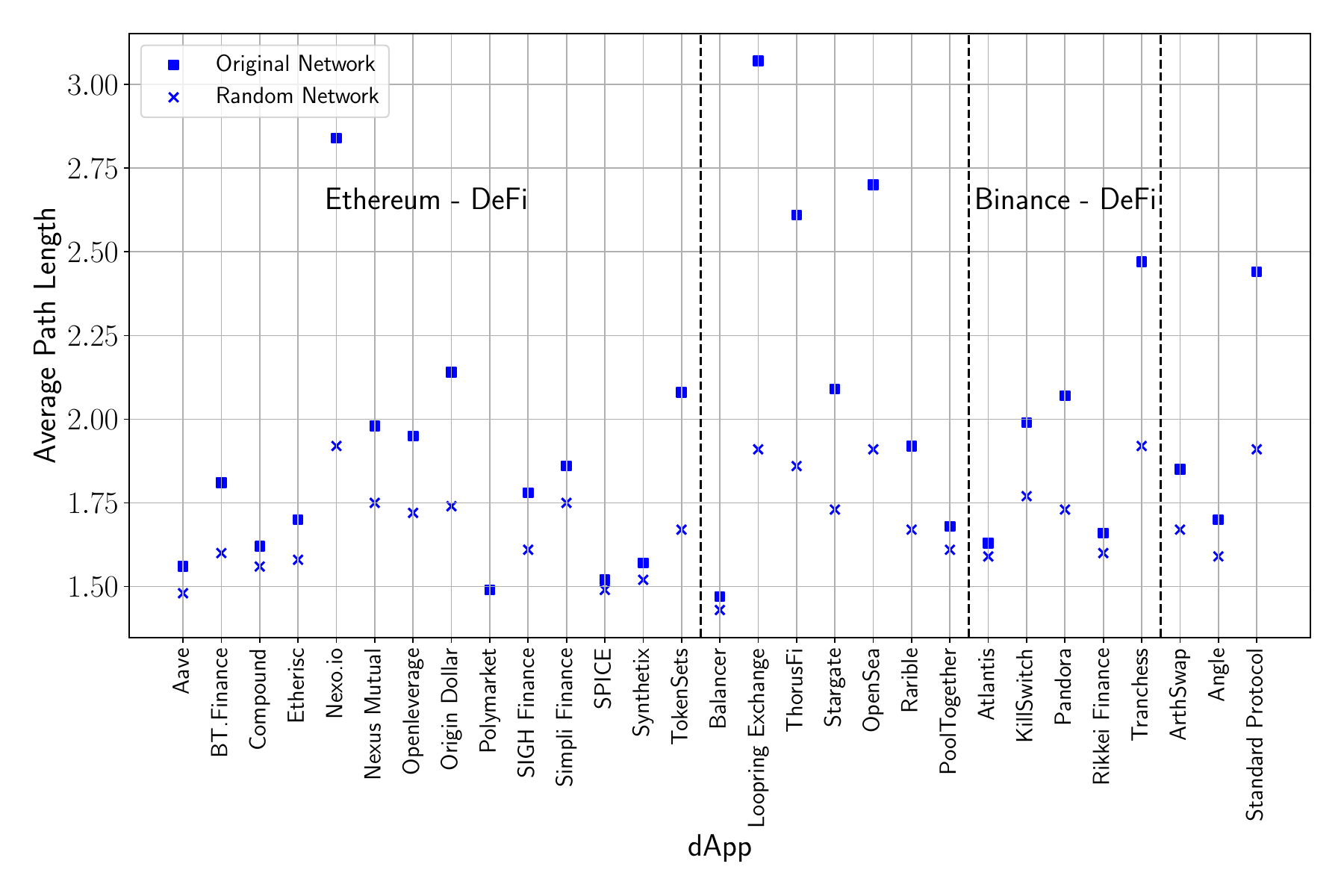}
    \caption{The plot shows a comparison between the average path length in the largest connected component of the real ($\small\blacksquare$) and random network ($\times$). We consider only the networks where the largest connected component contains more than $50$ nodes. }
    \label{fig:comparison path length}
\end{figure}

In network theory, small-world networks are distinguished from other networks by two properties: high clustering coefficient and short path lengths (as commonly observed in random networks). 
This network type is known for its ability to support rapid diffusion of information or processes across the network. Even if two nodes may be distant from each other, there are relatively short paths that indirectly link them, enabling quick transmission of information. 
In Fig. \ref{fig:plot 1} - bottom panel, we present the global clustering coefficients of nodes in the largest components of DeFi dApps. Similar results are obtained for non-DeFi related dApps and are shown in the Supplementary Material in Fig. \ref{fig:plot 2}.
In Fig. \ref{fig:comparison path length} we provide a comparison between the average path lengths in the largest connected components of the real networks and the ones of randomly generated networks with the same number of links and nodes. DApps exhibit high clustering coefficients and low average path lengths, similar to the ones of random networks with same number of nodes and links. The results suggest a similarity with the structure of a small world network, indicating the presence of substantial local interactions, efficient information flow within the component, and connectivity between functions, even when they are not directly linked, allowing fast information diffusion and effective collaboration among functions. 

\subsubsection*{Information diffusion}
\begin{figure}[H]
    \centering\includegraphics[width=0.7\textwidth]{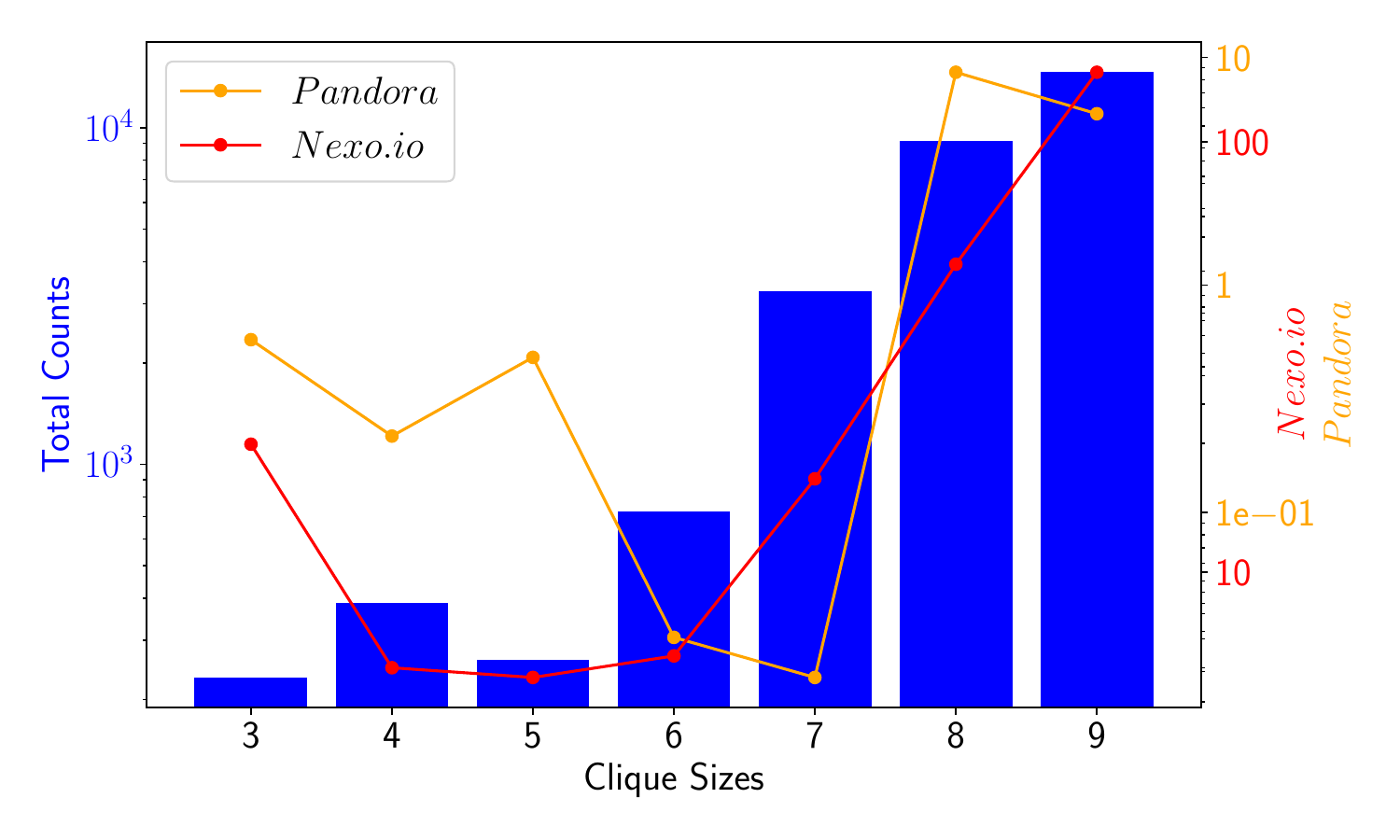}
    \caption{Maximal cliques' sizes across all networks. For each network, we analyse the list of the maximal cliques and normalize them by the size of the respective dApp. The cumulative cliques' sizes across all networks are shown in blue. There are two examples overlaying the bar chart: in orange, {\em Pandora} (Binance - DeFi, $40$ contracts), a medium-sized dApp, and in red, {\em Nexo.io} (Ethereum - DeFi, $176$ contracts), a large dApp.}
    \label{fig:maximal clique size}
\end{figure}

In graph theory, the concept of a clique is fundamental to understand the connectivity of networks. A clique is a group of vertices within a graph where each vertex is directly connected to every other vertex in the group, and its size is the number of vertices it contains. A maximal clique is a clique that cannot be extended by including one more adjacent vertex, meaning it is not a subset of a larger clique. Fig. \ref{fig:maximal clique size} shows the distribution of these maximal cliques of dimension $3$ to $9$ across the dApps networks, revealing a consistent trend across different sizes. The pattern shows the prevalence of large maximal cliques, meaning that the information on a function is just one step away from another, so the information diffusion process is immediate.
\subsection{Network Resilience} 

\begin{figure}[]
       \begin{minipage}{0.5\textwidth}
        \centering \includegraphics[width=\textwidth]{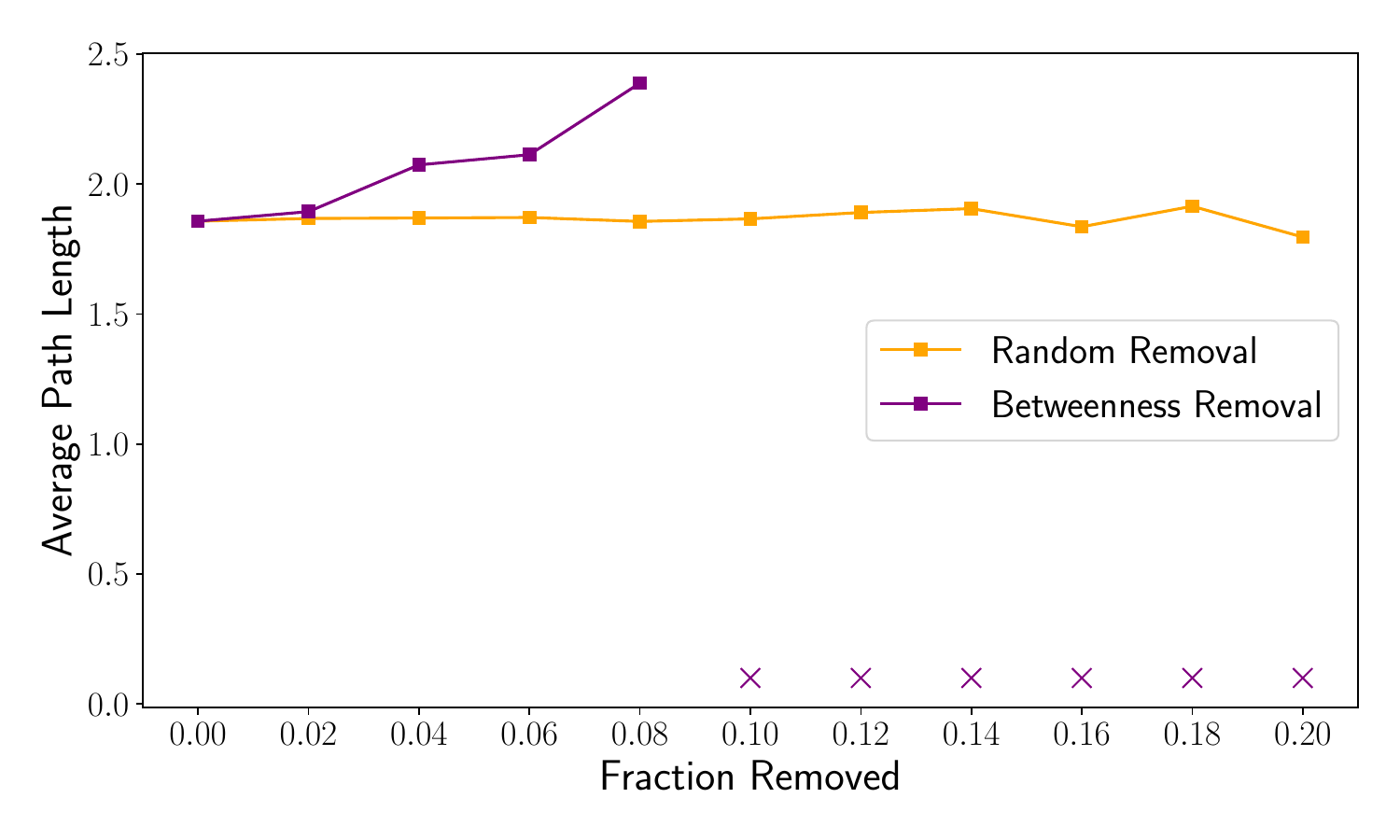} \subcaption {{\em Simpli Finance} (Ethereum - DeFi)}
    \end{minipage}
    \begin{minipage}{0.5\textwidth}
        \centering \includegraphics[width=\textwidth]{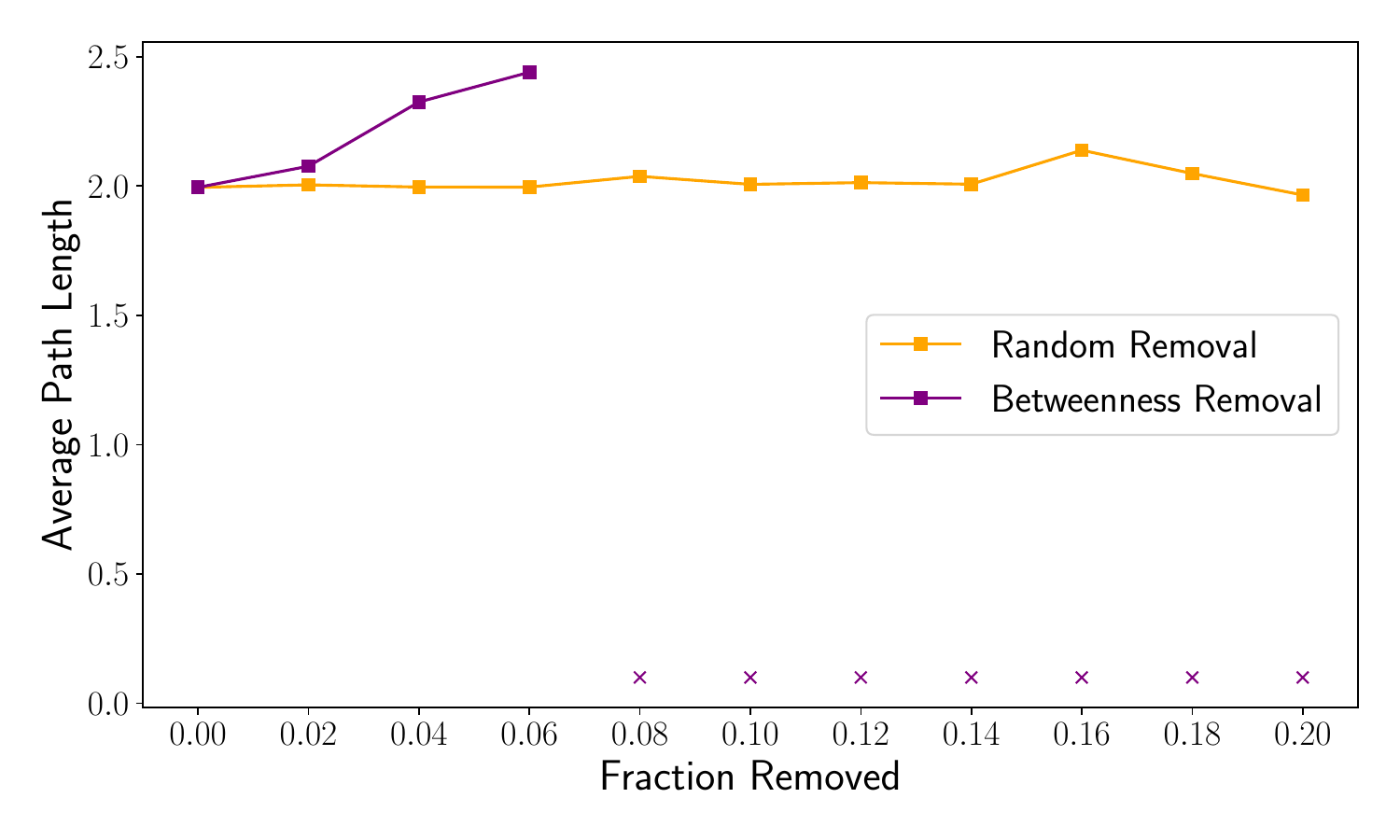} \subcaption{{\em KillSwitch} (Binance - DeFi)}
    \end{minipage}
    \caption{The plot shows the average path length within the largest connected component as a function of the fraction of nodes removed within the range of 0\% to 20\%. In orange, it represents the change in the average path length when nodes are randomly removed, while in purple, it illustrates the effect of removing nodes with the highest betweenness centrality, starting from the node with the highest betweenness and proceeding accordingly. The betweenness centrality calculation was performed at the beginning of the component analysis and it is not recomputed after each node removal. When the component becomes disconnected, as observed in the case of {\em Simpli Finance} when more than 8\% of nodes are removed, a cross is marked on the plot.} 
    \label{fig:avg path length removal}
\end{figure}

We can further analyse the largest connected components' resilience by examining the behaviour of average path length in the largest connected component in different conditions. In this analysis, we only consider networks with more than $50$ nodes in the largest component. In Fig. \ref{fig:avg path length removal}, we investigate how the average path length changes as an increasing fraction of nodes is removed, ranging from 0\% to 20\%. With targeted removal, the distances between the remaining nodes tend to increase, leading to the fragmentation into smaller, disconnected components. Targeting nodes with high betweenness centrality within dApps results in the complete disconnection of the component, a phenomenon not arising from the random node removal. The cases presented are the scenarios where the targeted removal of nodes keeps the component connected for as long as possible. However, in the majority of cases ($20$ dApps out of $29$), the component becomes disconnected after the removal of just 2\% of the nodes. In the event of a hacking attack, the dApps represented in the Fig. \ref{fig:avg path length removal} are the only ones that maintain uninterrupted information flow for a longer period, even when specific functions cease to operate. For all the other dApps, it is evident that an attack on a small percentage of those functions characterised by the highest betweenness centrality (indicating the  presence of significant information flow pathways), has the potential to disrupt the dApp's functionality. For instance, in the case of the dApp {\em Simpli Finance}, the component becomes disconnected when the functions {\em safeDecreaseAllowance} and {\em withdraw} are removed, respectively in charge of controlling the amount of tokens that can be withdrawn from an account and actually withdrawing them.
This critical threshold consistently appears to be around 2\% indicating that dApps are susceptible to potential targeted attacks. The results show a similar pattern when nodes are removed based on their degree centrality instead of betweenness centrality, indicating that the network's response to targeted removal is consistent across different centrality measures.
This result is also confirmed by the power-law degree distribution of nodes within the largest components, in Fig. \ref{fig:degree distrib} in the Supplementary Material. In a small world network with a degree distribution following a power-law, deletion of a random node rarely causes a dramatic increase in the average path length, because most shortest paths between nodes flow through supernodes, and if a peripheral node is deleted it is unlikely to interfere with passage between other peripheral nodes. As the fraction of peripheral nodes in a small world network is much higher than the fraction of supernodes, the probability of deleting an important node is very low.

\section{Discussion \& Conclusion}\label{sec:discussion}
We considered decentralised applications (dApps) of varying sizes, with different purposes, and deployed on various blockchain platforms, discovering consistent structural characteristics of contract and function networks across all of them. This consistency suggests that different development teams in the blockchain community adopt similar coding practices for smart contract design and development, regardless of the specific blockchain in use.  
In order to assess the resilience and security of dApps, we analysed the relationship within functions ad contracts. Ensuring that each function has a precise role when interfacing with a contract mitigates the effects of faults, but also facilitates a more systematic traceability, and verification of interactions for identifying anomalous behaviors. If modifications are required, all interactions of a given function can be redirected to an alternative contract, thus preserving the system's functionalities. In addition, by analysing and optimizing the frequency and nature of interactions between functions and their respective contracts, one can potentially minimize operational expenses like gas fees, consequently increasing the overall efficiency of the dApp.
The analysis of both contract and function networks within decentralized applications reveals interesting structural insights and interaction dynamics. 
\\
The networks of contracts interactions exhibit high sparsity, and a significant portion of the links consists of self-loops, suggesting that contracts within a dApp primarily interact with themselves. DApps are built with a focus on modular, self-sufficient contracts, forming distinct communities with limited external interactions and prioritizing security, fault isolation, and functional boundaries within dApps.
\\
We analyse the networks of functions interactions as well, in order to inspect code interactions within the dApp at a finer resolution. Contracts tend to distribute responsibilities among multiple functions, a practice that can be seen as a proactive measure against potential issues when relying solely on a single function to perform a complex task. Also the function networks maintain consistent characteristics that transcend the dApp's intended purpose. In this case there is a greater level of interconnection and a more intricate web of interactions among functions within the same dApp. DApps function networks exhibit a core largest component comprising functions interacting across multiple contracts, and encoding the core dApps' functionalities. Across all dApps, the organisation in distinct sub-components emerges, representing groups of functions with different - but secondary - tasks, defined within the same contract. Looking at the emergent structural organisation into core and secondary components, one may speculate that there is a coordinated and planned development of the core components, while the secondary contracts (interacting only with themselves) are disconnected parts added on an as-needed basis. An analysis of the timeline of development, monitoring code changes on Github, could shed further light on the architectural design and growth of the dApp infrastructure.

The largest component, containing interacting functions defined in different contracts, encodes the core functionality of the dApp. The core functions exhibit a high clustering coefficient and a low average shortest path length, resembling a small world model and suggesting significant local interactions, efficient information flow, and connectivity between functions. In the context of dApps, the small world structure implies that even when functions are not directly connected, efficient pathways for communication and interaction exist. Nodes with a high degree, which are closely correlated with nodes exhibiting high betweenness centrality (indicating their role as intermediaries between communities), as revealed by the degree distribution analysis, are relatively few: this means that a hacking attack on a random function does not significantly impact the overall dApp's functioning. However, if functions with high betweenness centrality are targeted in an attack, the largest component would immediately become disconnected and the information would stop flowing. This critical threshold consistently appears to be the 2\% of the total number of nodes. These findings emphasize that dApps are susceptible to potential targeted attacks, pointing to the importance of implementing robust strategies to mitigate potential vulnerabilities in these specific functions and guarantee their continued functionality.

Identifying the core areas and bridges within these networks allows us to monitor critical sections of the dApp, anticipate potential vulnerabilities, and explore ways to optimize computational costs. 
These patterns in the data are found through the examination of network structures. The subsequent phase should include the analysis of actual transaction data, monitoring the execution of on-chain code, and the actual usage of functions and contracts by dApps' users.

Moreover, further information and metrics to assess the characteristics of functions can be overlaid onto the network information, such as those extracted and analysed \cite{ibba2018preliminary}. This will allow further analysis of the quality of interactions among function calls. To comprehend the underlying reasons for the emergence of these patterns, it is essential to expand the information regarding the network structures with additional metrics, such as complexity costs or the number of lines of code in functions.

Indeed, not all vulnerabilities that exists are actually exploited \cite{perez}, but they still constitute a potential threat to the normal functioning of the platform. The exploitation of vulnerabilities in targeted attacks, leads to a decrease in trust in decentralised platforms, hindering users' adoption \cite{Auer2023} and institutional investors' support \cite{Mungo}.

\newpage
\subsection*{Acknowledgement} S.B., G.D., R.N. and M.O. acknowledge support from the Ethereum foundation grant FY23-1048.
%{\small
%\printbibliography}
\bibliographystyle{unsrt}

\newpage
\appendix

\section{List of dApps}\label{app:list of dapps}
The categorisation of dApps used in this work was done following the classification proposed by 
\href{https://www.bitdegree.org/crypto-tracker/top-dapps}{BitDegree} and \href{https://dappradar.com}{DappRadar}.

\begin{table}[H]
\centering
\begin{tabular}{|c|c|c|}
\hline
\textbf{Blockchain} & \textbf{Category} & \textbf{dApps}                                                                                                                                                                                                                                                                                                                \\ \hline
Ethereum            & Collectibles      & Async Art; Audius; Cryptovoxels                                                                                                                                                                                                                                                                                               \\ \hline
Ethereum            & DeFi              & \begin{tabular}[c]{@{}c@{}}1inch Network; Aave; BondAppetit; BT.Finance$^*$; \\ Compound; Etherisc; Naos Finance; Nexo.io;\\ Nexus Mutual; Openleverage$^*$; Origin Dollar; Polymarket; \\ PWN; Rari Capital; Rocket Pool; SIGH Finance; \\ Simpli Finance$^*$; SPICE; SWAPP Protocol$^*$; Synthetix$^*$; \\ TokenSets; Tsunami; UMA$^*$\end{tabular} \\ \hline
Ethereum            & Exchanges         & \begin{tabular}[c]{@{}c@{}}Balancer; Brickblock; Loopring Exchange; Plexus$^*$;\\ Popsicle V3 Optimizer$^*$; ThorusFi$^*$; Uniswap\end{tabular}                                                                                                                                                                                           \\ \hline
Ethereum            & Gambling          & DSG; Stargate                                                                                                                                                                                                                                                                                                                 \\ \hline
Ethereum            & Games             & Axie Infinity; DARK FOREST; Gods Unchained; Marble.Cards                                                                                                                                                                                                                                                                      \\ \hline
Ethereum            & High-risk         & Proof of Fair Launch                                                                                                                                                                                                                                                                                                          \\ \hline
Ethereum            & Marketplaces      & Foundation; Fractional; OpenSea; Rarible; SuperRare                                                                                                                                                                                                                                                                           \\ \hline
Ethereum            & Other             & \begin{tabular}[c]{@{}c@{}}Aragon Fundraising; AZTEC; Ethereum Name Service; \\ Polymath; PoolTogether; Tornado Cash\end{tabular}                                                                                                                                                                                             \\ \hline
Binance Smart Chain & DeFi              & \begin{tabular}[c]{@{}c@{}}Atlantis; BabySwap; FarmHero; KillSwitch;\\ Pandora; Rikkei Finance; Tranchess; Venus\end{tabular}                                                                                                                                                                                                 \\ \hline
Binance Smart Chain & Gambling          & LuckyChip                                                                                                                                                                                                                                                                                                                     \\ \hline
Binance Smart Chain & Gaming            & NomadLand                                                                                                                                                                                                                                                                                                                     \\ \hline
Astar               & DeFi              & ArthSwap                                                                                                                                                                                                                                                                                                                      \\ \hline
Polygon             & DeFi              & Angle                                                                                                                                                                                                                                                                                                                         \\ \hline
Polygon             & Gambling          & Reality Cards                                                                                                                                                                                                                                                                                                                 \\ \hline
Shiden              & DeFi              & Standard Protocol                                                                                                                                                                                                                                                                                                             \\ \hline
ThunderCore         & DeFi              & Staking Pool                                                                                                                                                                                                                                                                                                                  \\ \hline
\end{tabular}
\end{table}
 
$(^*)$: cross-chains dApps \\
Popsicle V3 Optimizer: Ethereum, Avalanche, Fantom, Binance, Polygon; \\
BT.Finance: Ethereum, Binance Smart Chain; \\
Simpli Finance: multichain; \\
ThorusFi: multichain; \\
SWAPP Protocol: Ethereum, Binance Smart Chain; \\
Openleverage: Ethereum, Binance Smart Chain; \\
Plexus: Ethereum, Binance Smart Chain, Optimism, Polygon; \\
Synthetix: Ethereum, Optimism; \\
UMA: Ethereum, Optimism.
\section{Contracts Networks}\label{app:contracts networks}
\begin{figure}[H]
    \centering
    \includegraphics[width=0.44\textwidth]{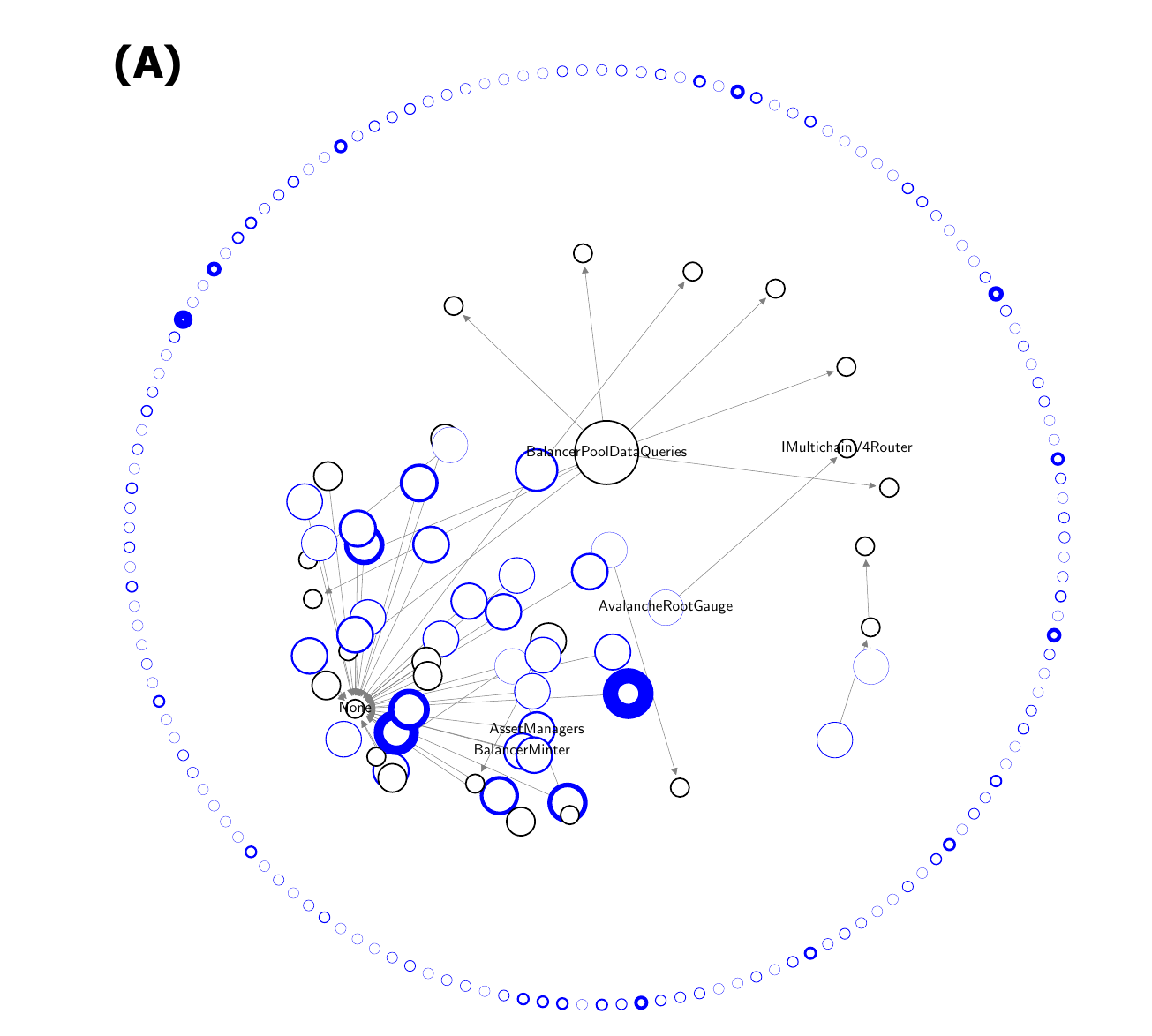}
      \includegraphics[width=0.42\textwidth]{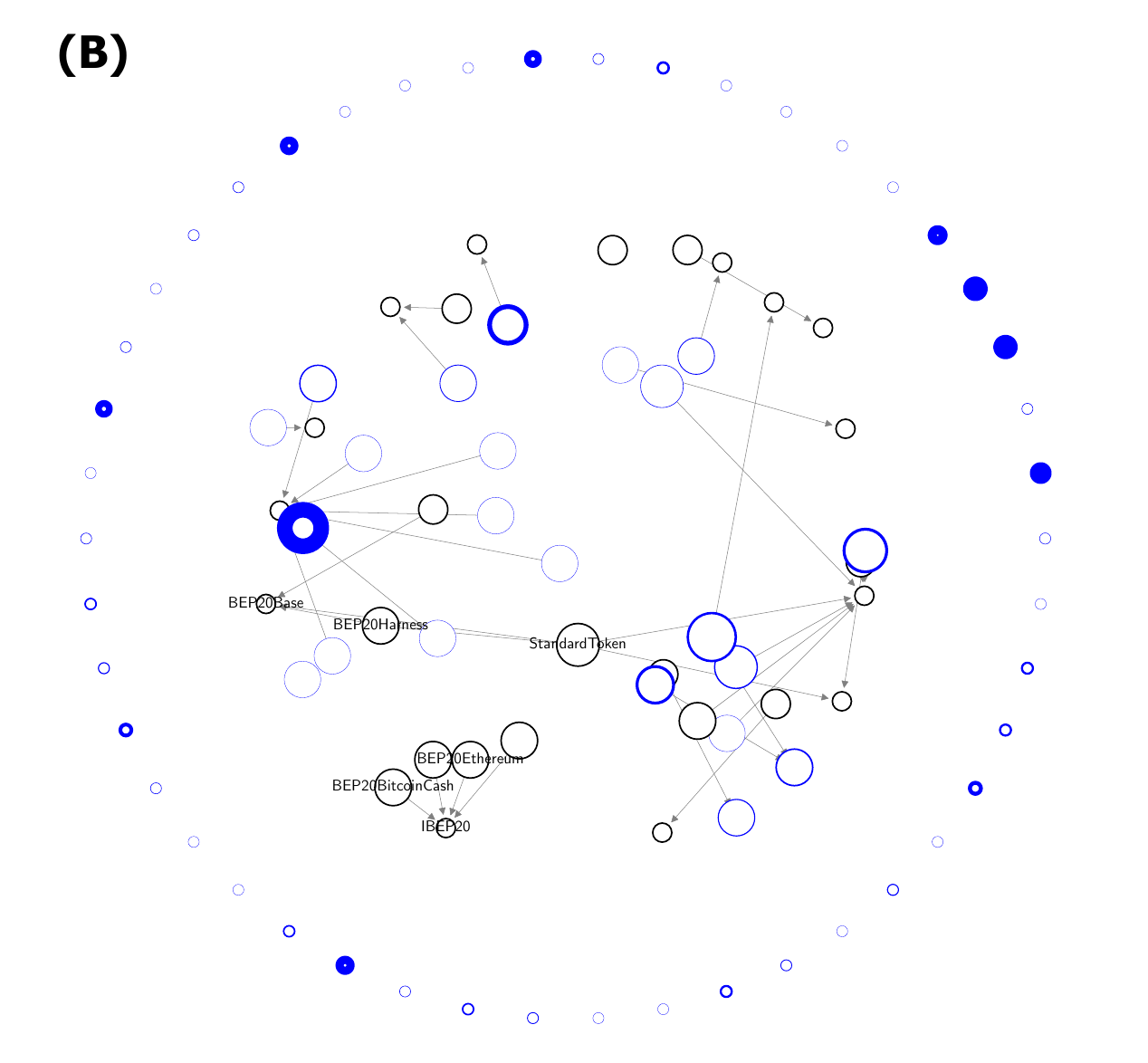}\\
        \includegraphics[width=0.44\textwidth]{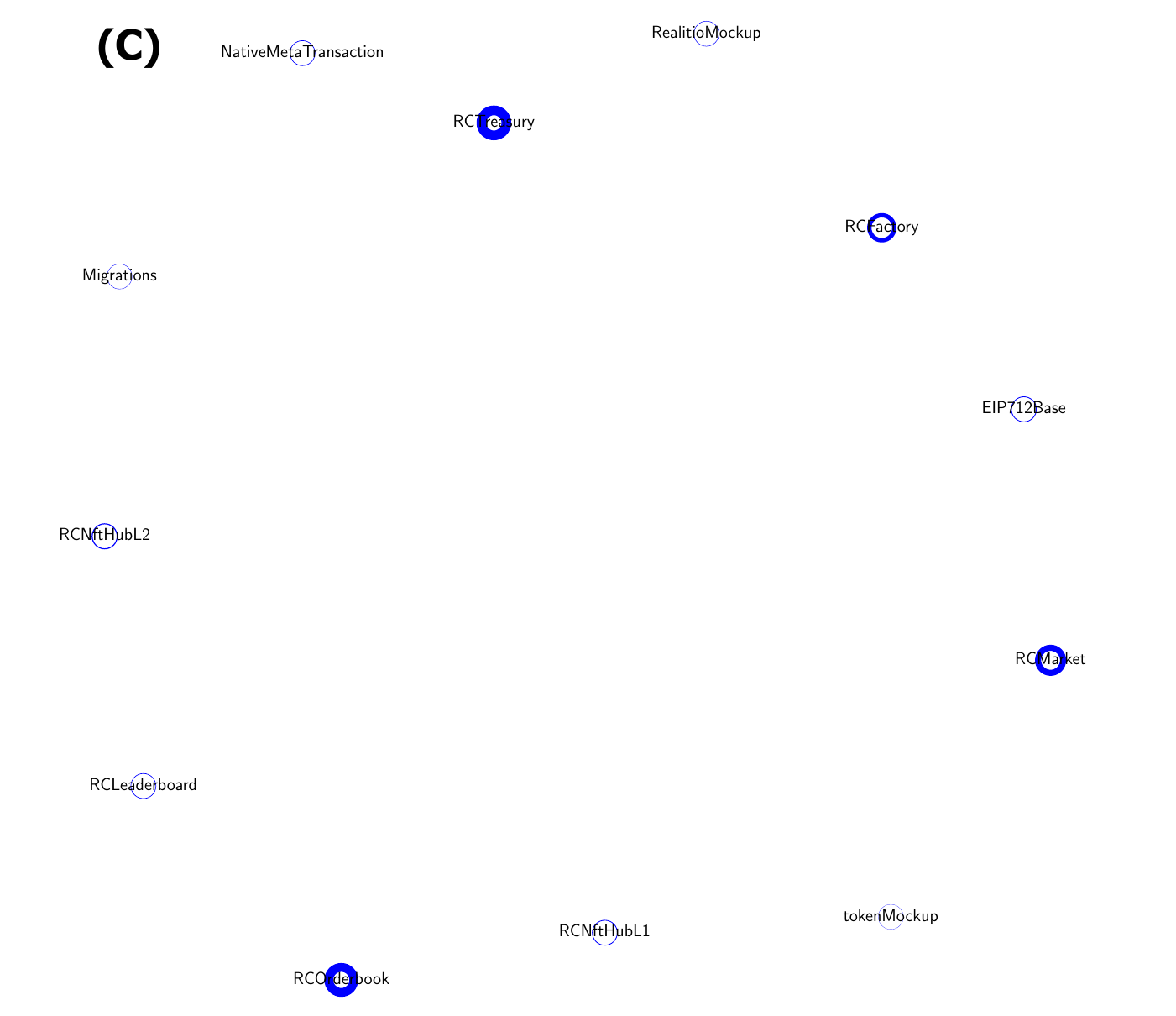}
    \caption{(A) {\em Balancer} Contracts Network (Ethereum - Exchanges).  (B) {\em Venus} Contracts Network (BNB - DeFi).  (C) {\em Reality Cards} Contracts Network (Polygon - Gambling). In blue, the nodes with self-loops. The width of the border of each blue node corresponds to the weight of its self-loop link. The names of the contracts with the highest betweenness centrality are listed.}
    \label{fig:contracts}
\end{figure}

In this section, we show further examples of dApp contracts networks. In Fig. \ref{fig:contracts}, panel (A), we show {\em Balancer} Contracts Network,  a dApp deployed in Ethereum and belonging to the category exchanges. It is a large dApp consisting of $193$ contracts. The total balance is $\$223.7T $ and it is ranked $\#33$ in the category Exchanges, and $\#359$ in the General category in dApp radar. In blue, we highlight the nodes with self-loops, where the width of the border of each blue node corresponds to the weight of its self-loop link. The names of the contracts with the highest betweenness centrality are listed. In Fig. \ref{fig:contracts}, panel (B), we show {\em Venus} Contracts Network, a dApp deployed on Binance (BNB) and belonging to the DeFi category. It is classified as a large dApp with $83$ contracts and a total balance of $\$574.32B $. It is ranked $\#5$ in DeFi and $\#10$ in General. In Panel (C), we present {\em Reality Cards} Contracts Network, deployed on the Polygon blockchain network and listed in the Gambling category. It is a rather small dApp with $12$ contracts and a balance of $\$13,32k$.  It is ranked $\#1221$ in the Gambling category.
\section{Functions Networks}\label{app:functions networks}
\begin{figure}[H]
    \begin{minipage}{0.47\textwidth}
        \centering \includegraphics[width=\textwidth]{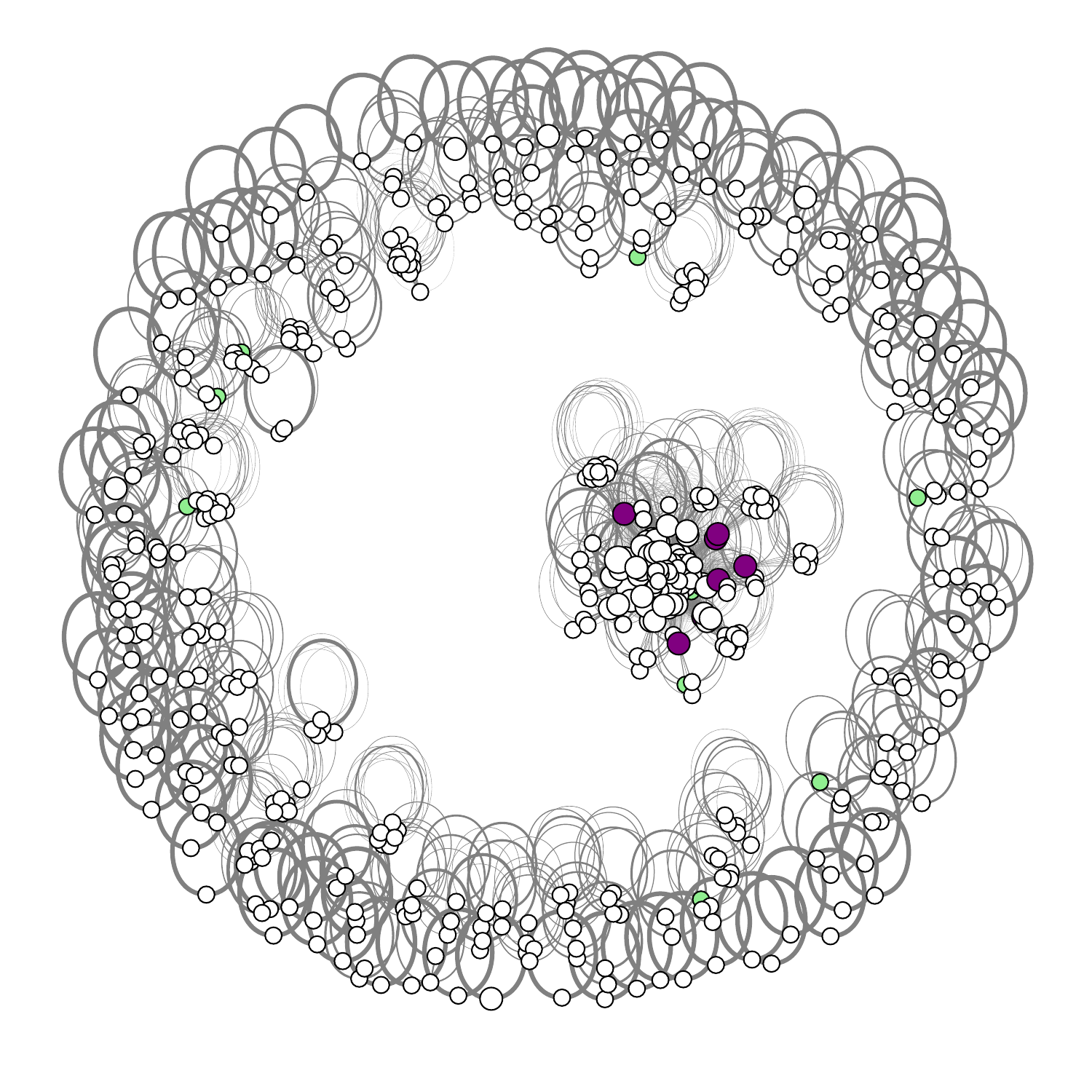} \subcaption{Network for functions' interactions pre-filter}
    \end{minipage}%
    \begin{minipage}{0.47\textwidth}
        \centering \includegraphics[width=\textwidth]{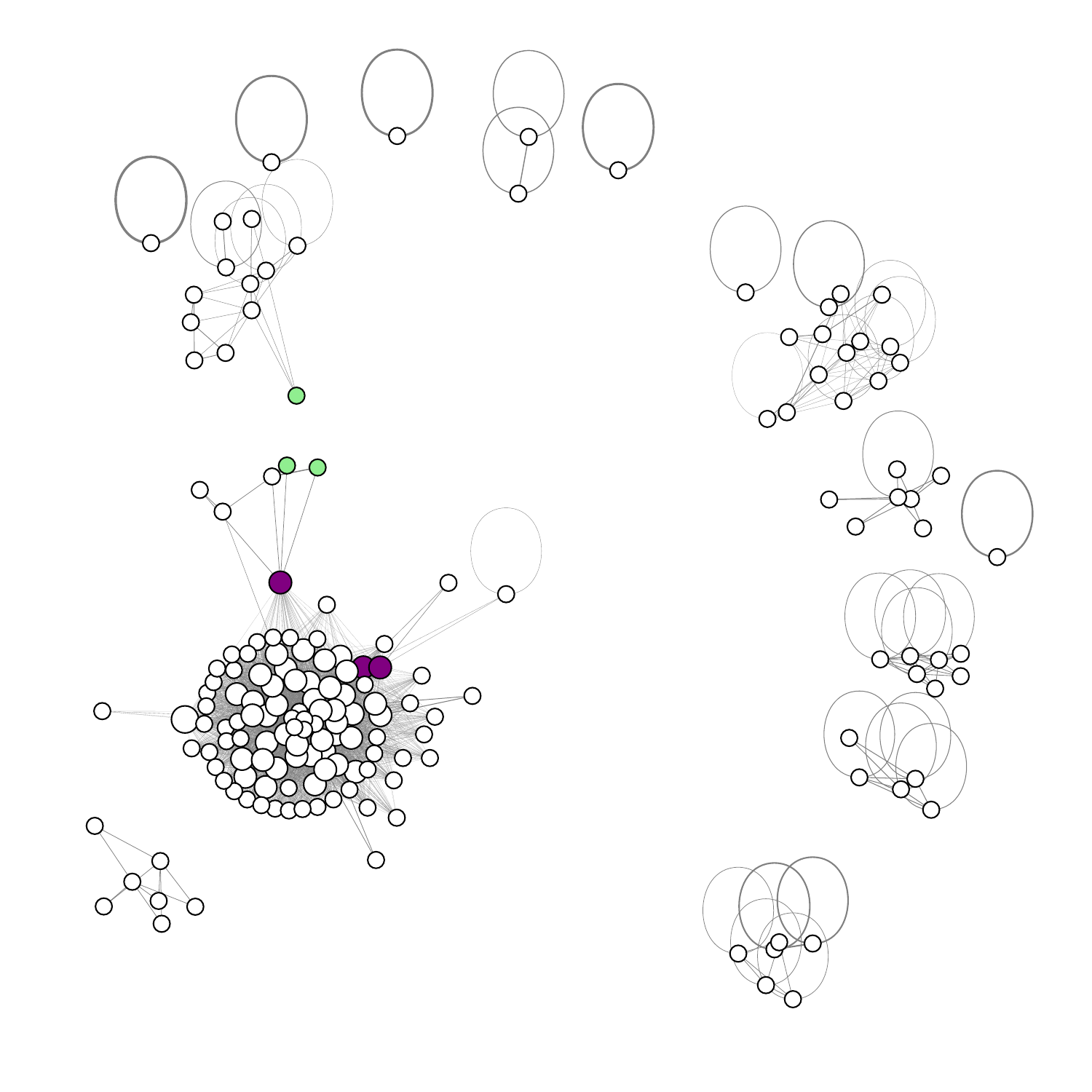} \subcaption{Network for functions' interactions post-filter}
    \end{minipage}
    \caption{{\em Balancer} Functions Network (Ethereum - Exchanges). In purple the nodes with highest betweenness (i.e. {\em \_addGauges}, {\em removeToken}, {\em removeAllowedAddress}), in green the nodes with highest clustering coefficient (i.e. {\em \_returnLeftoverEthIfAny}, {\em addGaugesWithVerifiedType}, {\em \_onSwapMinimal}). The purple functions are in charge of claiming the measure of liquidity provided by users or removing an address that was previously authorised for given transactions from the pool. The green functions return the ETHs that advance as a result of a given transaction, retrieve the status of the pool and verify that it is active.
    }\label{fig:balancer funzioni}
\end{figure}

\begin{figure}[H]
    \begin{minipage}{0.47\textwidth}
        \centering \includegraphics[width=\textwidth]{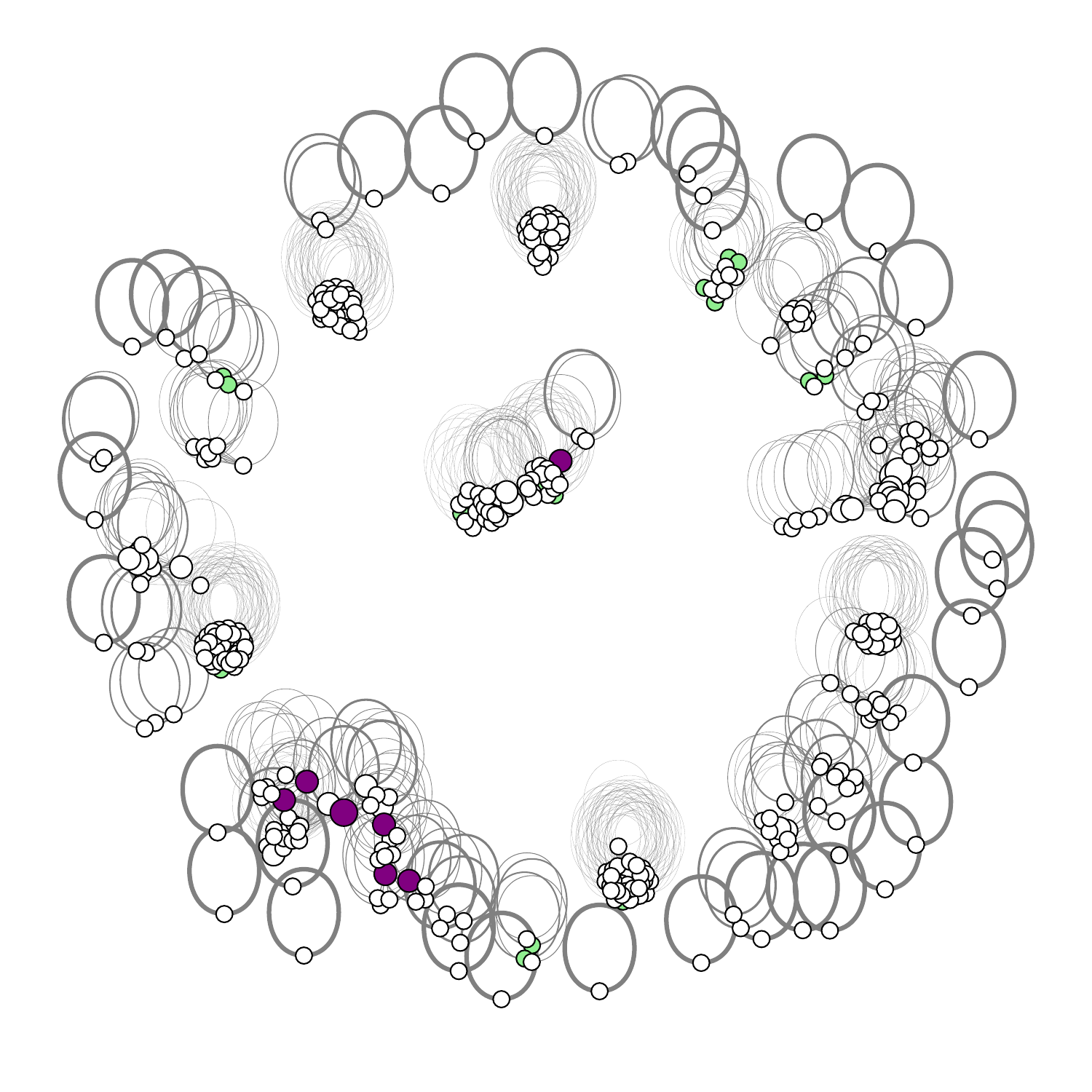} \subcaption{Network for functions' interactions pre-filter}
    \end{minipage}%
    \begin{minipage}{0.47\textwidth}
        \centering \includegraphics[width=\textwidth]{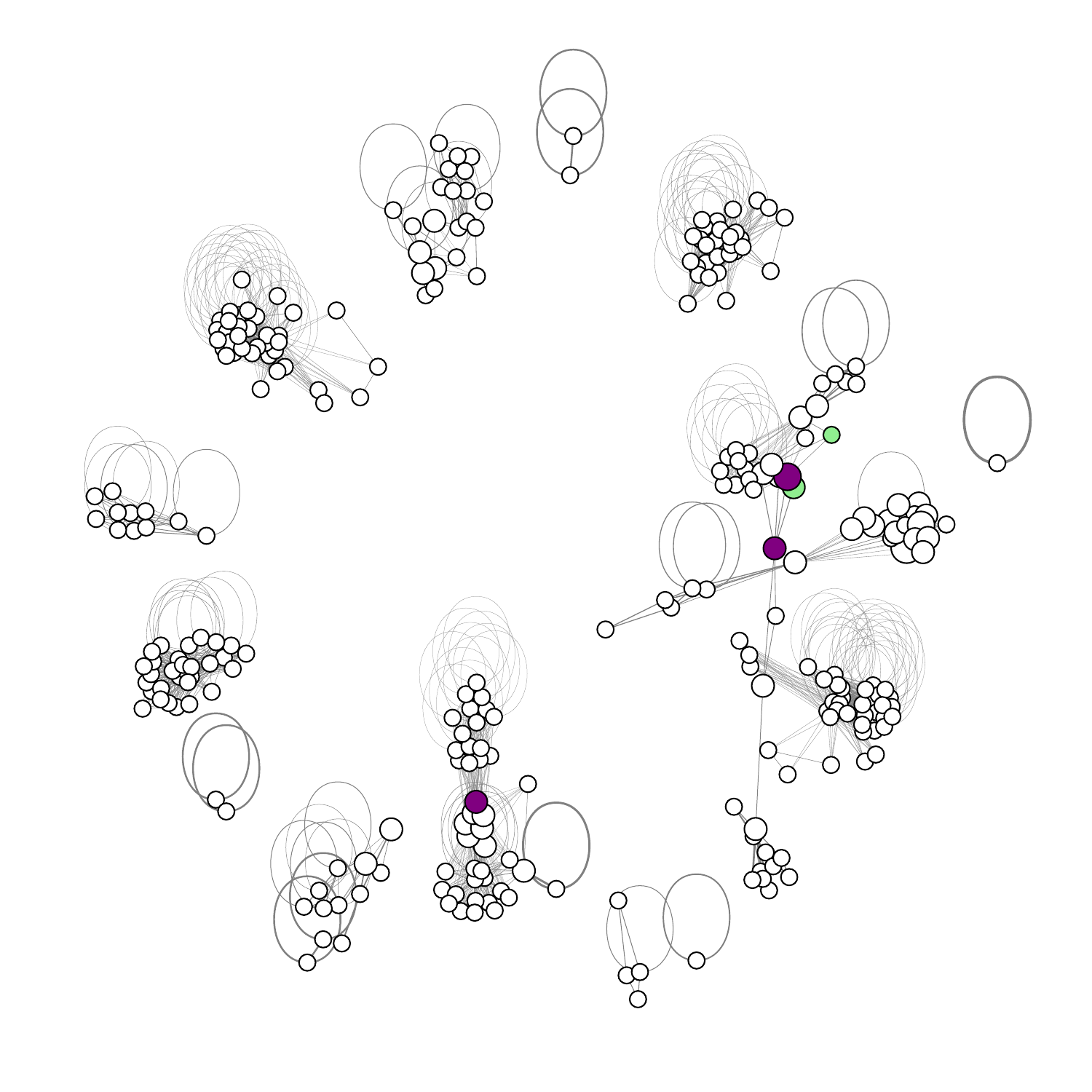} \subcaption{Network for functions' interactions post-filter}
    \end{minipage}
    \caption{{\em Venus} Functions Network (BNB - DeFi). In purple the nodes with highest betweenness (i.e. {\em reclaimToken}, {\em transferFrom}, {\em approve}), in green the nodes with highest clustering coefficient (i.e. {\em \_burn}, {\em allocateTo}, {\em \_transferOwnership}). The purple functions are in charge of transferring or claiming an amount of tokens from one address to another and confirming that it has been done successfully. The green functions destroy an amount of tokens relating to an account or transfer their ownership.
    }\label{fig:venus funzioni}
\end{figure}

\begin{figure}[H]
    \begin{minipage}{0.47\textwidth}
        \centering \includegraphics[width=\textwidth]{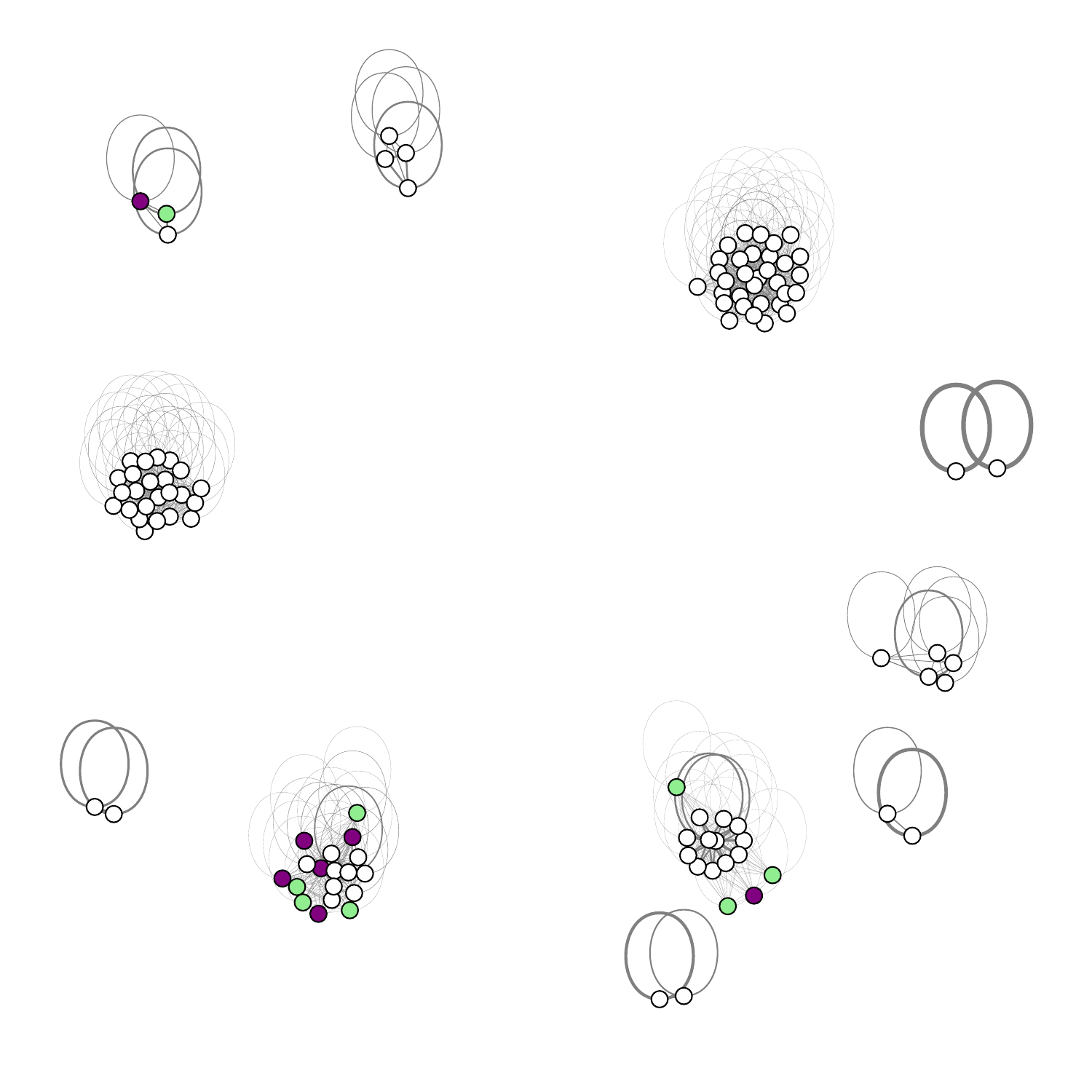} \subcaption{Network for functions' interactions pre-filter}
    \end{minipage}%
    \begin{minipage}{0.47\textwidth}
        \centering \includegraphics[width=\textwidth]{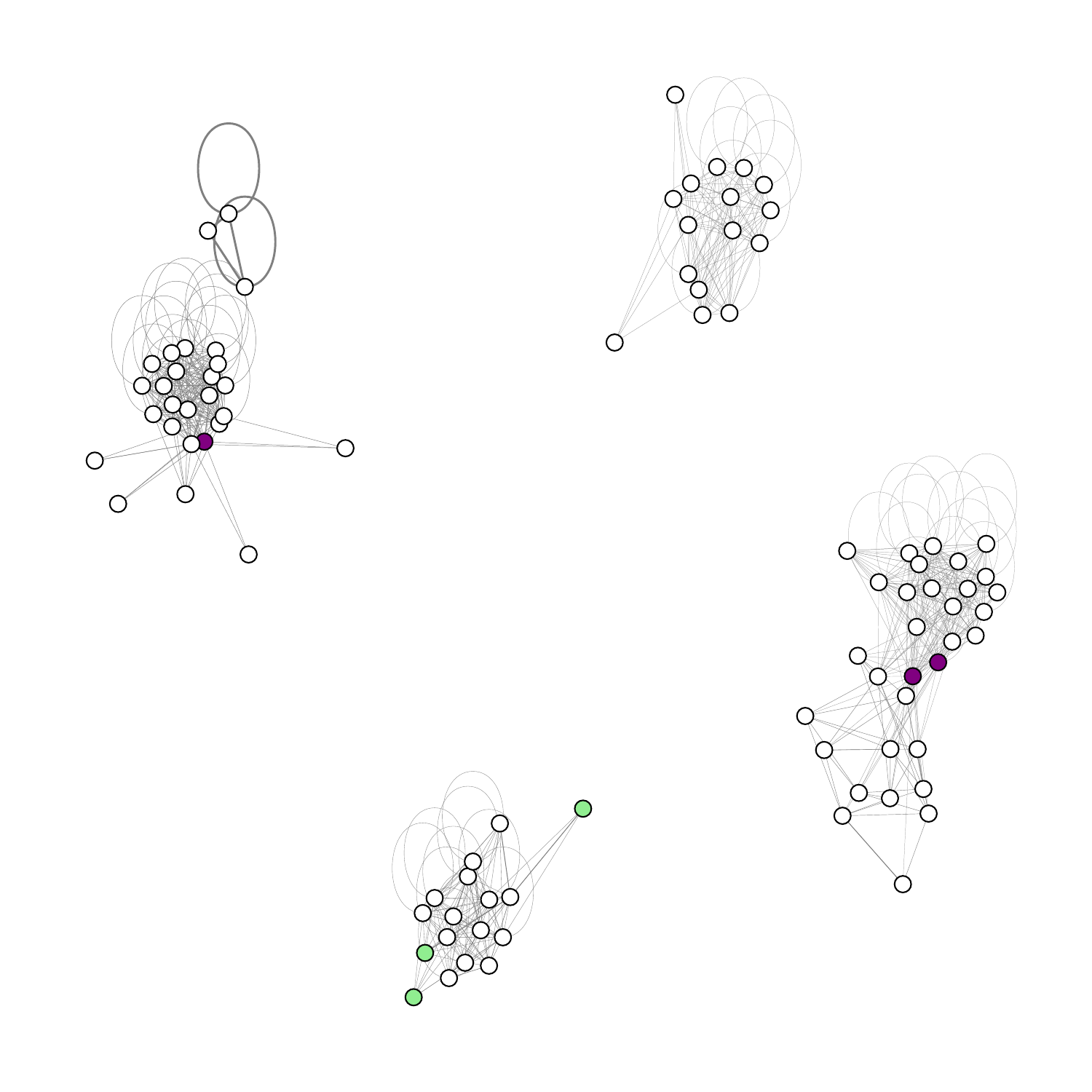} \subcaption{Network for functions' interactions post-filter}
    \end{minipage}
    \caption{{\em Reality Cards} Functions Network (Polygon - Gambling). In purple the nodes with highest betweenness (i.e. {\em addMarket}, {\em decreaseBidRate}, {\em changeApprovedAffiliatesOnly}), in green the nodes with highest clustering coefficient (i.e. {\em \_postQuestionToOracle}, {\em claimCard}, {\em exit}). The purple functions are in charge of removing the bet based on its withdrawal and changing the status of a market affiliate. The green functions require a card during the game and end the current game.
    }\label{fig:reality cards funzioni}
\end{figure}

In this section, we show further examples of dApp function networks before and after applying the filter. We consider as in Sec. \ref{app:contracts networks} the following dApps: {\em Balancer} (Fig. \ref{fig:balancer funzioni}), {\em Venus} (Fig. \ref{fig:venus funzioni}), and {\em Reality Cards} (Fig. \ref{fig:reality cards funzioni}). We also highlight key functions and their role in the dApp within the networks.

\section{Further results} \label{app:res}
\begin{figure}[H]
        \centering \includegraphics[width=0.7\textwidth]{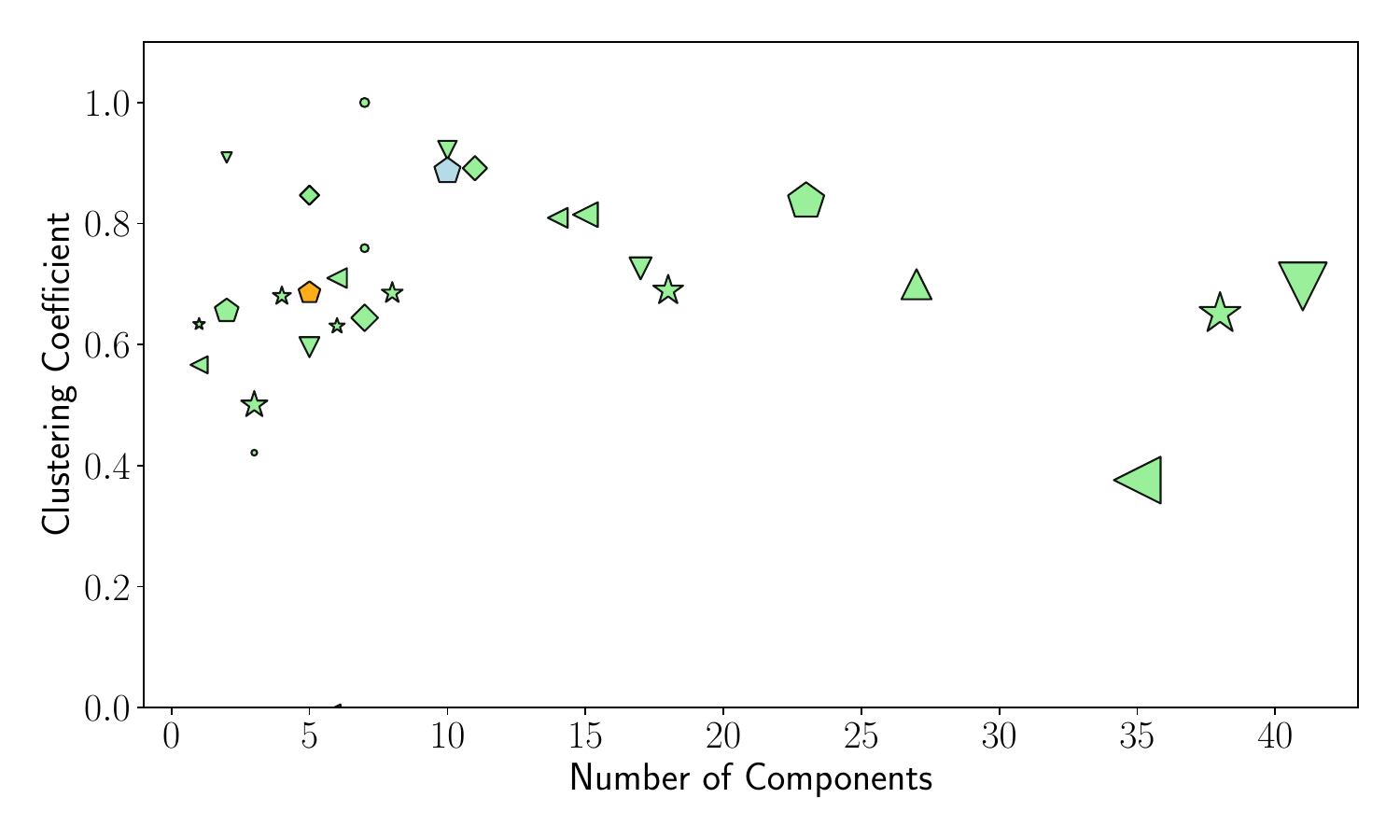}
    \caption{Scatter plot of the global clustering coefficient in the largest components vs the number of components. Each dot represents a dApp, and the size of the dot is proportional to the number of functions in the specific dApp. Legend: Collectibles ($\bullet$); Exchanges ($\bigstar$); Gambling:  ($\pentagofill$); Games ($\blacklozenge$); High-risk ($\blacktriangle$); Marketplaces: ($\blacktriangledown$); Other: ($\blacktriangleleft$).}
    \label{fig:plot 2}
\end{figure}

\begin{figure}[H]
    \centering
    \includegraphics[width=0.7\textwidth]{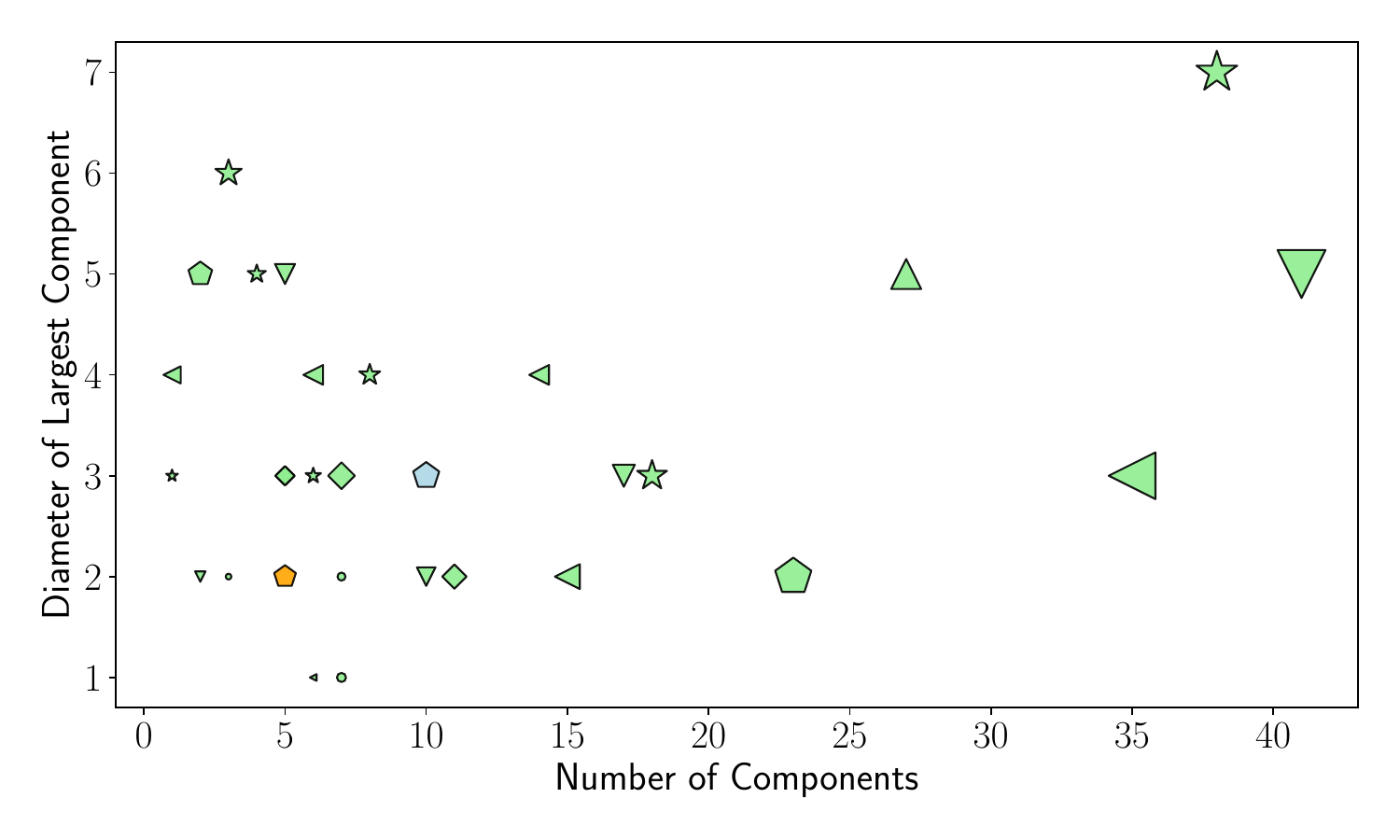}
    \caption{Scatter plot of the diameter of the largest connected component vs the number of components, for non-DeFi related dApps. Each symbol represents a dApp, and its size is proportional to the number of functions in the specific dApp. Legend: Collectibles ($\bullet$); Exchanges ($\bigstar$); Gambling:  ($\pentagofill$); Games ($\blacklozenge$); High-risk ($\blacktriangle$); Marketplaces: ($\blacktriangledown$); Other: ($\blacktriangleleft$).}
    \label{fig:plot 3}
\end{figure}

In this section, we include further analysis on cluster coefficients and diameter of the largest connected component for non-DeFi related dApps (see Figs. \ref{fig:plot 2}, \ref{fig:plot 3}). Patterns similar to those observed by restricted the set to DeFi only dApps are present (see Fig. \ref{fig:plot 1}).

\begin{figure}[H]
    \centering
    \includegraphics[width=0.7\textwidth]{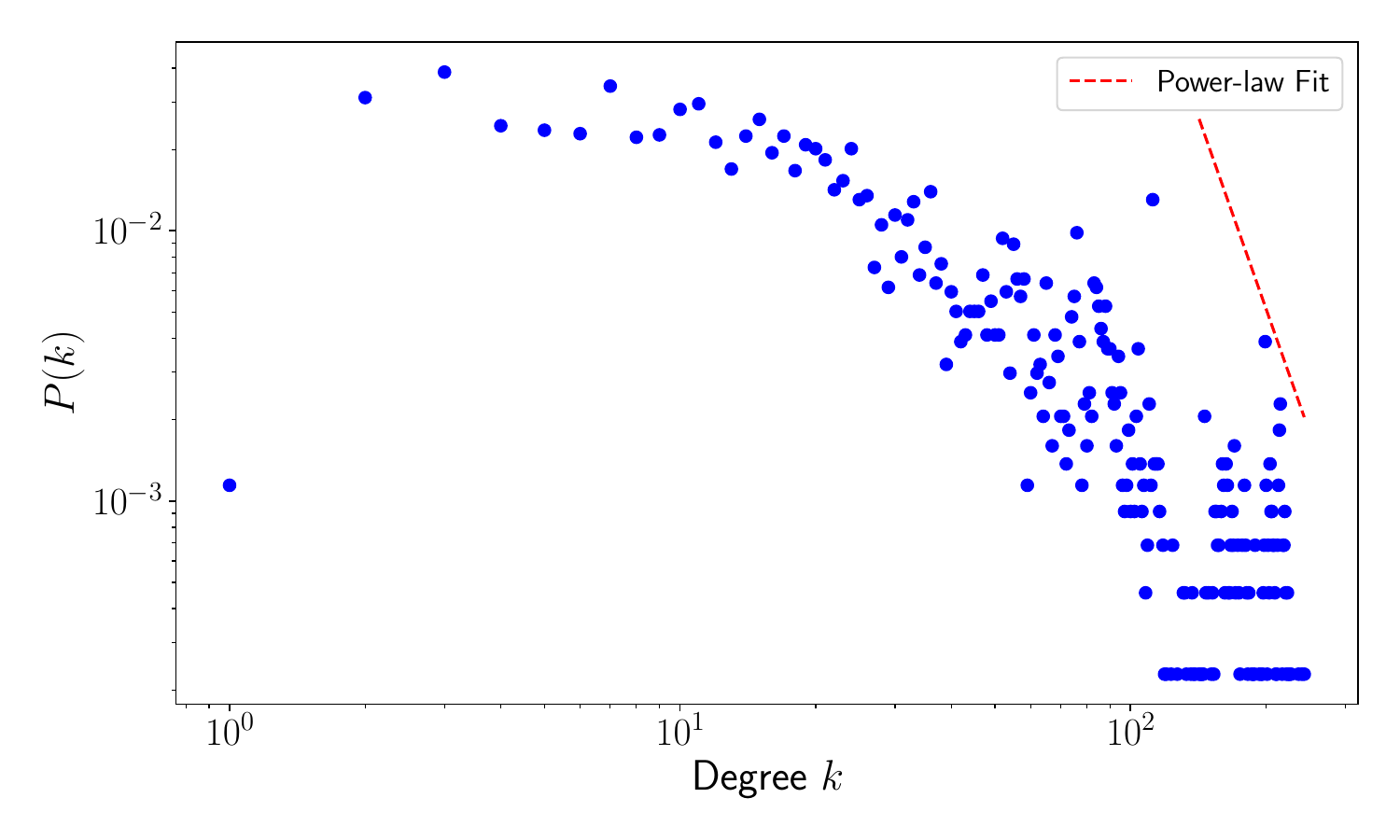} 
    \caption{In blue, the degree distribution of the largest components is displayed. In red, a power-law fit of the distribution, with $\alpha = 4.7$ and $x_{min} = 142$.}
    \label{fig:degree distrib}
\end{figure}

In Fig. \ref{fig:degree distrib} we show the degree distribution computed across all nodes in all dApps. The power-law degree distribution suggests that most nodes have few neighbors, while some supernodes (closely correlated with the nodes with highest betweenness) have a higher number of neighbors.

\end{document}